\documentclass[aps,prx,reprint,floatfix,amsmath,amssymb,superscriptaddress]{revtex4-2}

\usepackage{graphicx}
\graphicspath{{PRX_final/}{./}}
\usepackage{amssymb}
\usepackage{mathtools}
\usepackage{microtype}
\usepackage{physics}
\usepackage{bm}
\usepackage{mathrsfs}
\usepackage{booktabs}

\usepackage{xcolor}

\usepackage{hyperref}
\usepackage{bookmark}

\hypersetup{
  pdftitle={Analytic detector responses and entanglement in constant-curvature spacetimes},
  pdfauthor={Dyuman Bhattacharya, Jiayue Yang, Ming Zhang, and Robert B. Mann},
  colorlinks=true,
  linkcolor=black,
  citecolor=black,
  urlcolor=black
}

\makeatletter
\let\Hy@raisedlink\@firstofone
\makeatother

\DeclareMathOperator{\erfc}{erfc}
\DeclareMathOperator{\erfi}{erfi}
\DeclareMathOperator{\sgn}{sgn}

\allowdisplaybreaks[2]
\begin{document}

\title{Analytic detector responses and entanglement in constant-curvature spacetimes}

\author{Dyuman Bhattacharya}
\email{d7bhatta@uwaterloo.ca}
\affiliation{Department of Physics and Astronomy, University of Waterloo, Waterloo, Ontario N2L 3G1, Canada}

\author{Jiayue Yang}
\email{j43yang@uwaterloo.ca}
\affiliation{Department of Applied Mathematics, University of Waterloo, Waterloo, Ontario N2L 3G1, Canada}
\affiliation{Institute for Quantum Computing, University of Waterloo, Waterloo, Ontario N2L 3G1, Canada}
\affiliation{Perimeter Institute for Theoretical Physics, 31 Caroline St. N., Waterloo, Ontario N2L 2Y5, Canada}
\affiliation{Department of Physics and Astronomy, University of Waterloo, Waterloo, Ontario N2L 3G1, Canada}

\author{Ming Zhang}
\email{m539zhan@uwaterloo.ca}
\affiliation{Department of Physics and Astronomy, University of Waterloo, Waterloo, Ontario N2L 3G1, Canada}
\affiliation{Perimeter Institute for Theoretical Physics, 31 Caroline St. N., Waterloo, Ontario N2L 2Y5, Canada}
\affiliation{Department of Physics, Jiangxi Normal University, Nanchang 330022, China}

\author{Robert B.~Mann}
\email{rbmann@uwaterloo.ca}
\affiliation{Department of Applied Mathematics, University of Waterloo, Waterloo, Ontario N2L 3G1, Canada}
\affiliation{Institute for Quantum Computing, University of Waterloo, Waterloo, Ontario N2L 3G1, Canada}
\affiliation{Perimeter Institute for Theoretical Physics, 31 Caroline St. N., Waterloo, Ontario N2L 2Y5, Canada}
\affiliation{Department of Physics and Astronomy, University of Waterloo, Waterloo, Ontario N2L 3G1, Canada}

\begin{abstract}
We derive analytic response functions for Unruh--DeWitt detectors coupled to real, massless, conformally coupled scalar fields, to distinguish the effects of the field spectrum and global geometry from those of the interaction protocol. For Gaussian-switched static pairs in Minkowski, anti-de Sitter, and de Sitter spacetimes, we obtain the leading-order excitation probabilities and two-detector coherences in spacetime dimension $\mathcal D=d+1\ge3$. Time-ordered coherence requires distinct spatial worldlines; the local response and single-excitation coherence also admit coincidence. For equal-redshift pairs, Gaussian switching separates the gap dependence of the time-ordered coherence. In de Sitter space, KMS detailed balance fixes the gap that maximizes the ratio of nonlocal coherence to local excitation, giving a single-gap test for the existence of leading-order entanglement and restricting any entangling gaps to one bounded interval. When entanglement is present, the concurrence itself peaks at a smaller positive gap. In the three-dimensional BTZ black hole, an integrated Legendre series gives the sharply switched response of a radially infalling detector. A uniform image-sum estimate establishes logarithmic growth near the singularity and determines its coefficient from the covering-AdS response. A sufficiently smooth monotone onset removes the finite-time onset glitches but leaves a positive logarithmic coefficient; the accumulated leading-order response remains finite. Independent numerical calculations test the analytic expressions, their convergence properties, and these physical distinctions.
\end{abstract}
\maketitle

\section{Introduction}
\label{sec:introduction}

Relativistic quantum information models localized probes as two-level quantum mechanical systems known as Unruh--DeWitt (UDW) detectors that couple to quantum fields \cite{Unruh1979evaporation,DeWitt1979,Smith.2017}. Their excitation probabilities and correlations are determined by the spacetime geometry, detector trajectories, switching, and field state. Entanglement harvesting provides a representative application: initially separable detectors can become entangled through local field interactions \cite{PozasKerstjens:2015}. Extensions include tripartite entanglement \cite{MendezAvalos2022} and magic-resource harvesting \cite{Nystrom2025Magic}.

Calculations of detector responses and correlations in relativistic quantum information usually rely on the traditional method of numerical integration. Notwithstanding the generality of   such methods, they can be computationally costly and may also obscure the physical origin of qualitative features in RQI phenomena. Indeed, detector responses and correlations generally depend simultaneously on the quantum field, detector trajectory, switching protocol, and spacetime geometry, making these different contributions difficult to disentangle from numerical results. Analytic control is therefore valuable not only for computational efficiency, but also for exposing the underlying spectral and geometric structure, making explicit the dependence on the physical parameters of the problem, and identifying qualitative and universal features that may be difficult to recognize from numerical results alone. Constant-curvature spacetimes provide a particularly useful setting for this purpose: their two-point functions possess substantial analytic structure, while still exhibiting nontrivial redshift, boundary, thermal, and global effects. This motivates us to develop analytic expressions for the detector response functions and reduced density-matrix elements relevant to RQI in these spacetimes.

Specifically, 
at leading order, the field two-point function along the worldlines fixes the detector state \cite{Schlicht_2004}. For example, for  two detectors $A$ and $B$ initially in their ground states, let $P_A,P_B$ be the excitation probabilities and $X$ the coherence between $|00\rangle$ and $|11\rangle$. Entanglement harvesting requires $|X|>\sqrt{P_AP_B}$ \cite{PozasKerstjens:2015}. We evaluate this inequality at finite Gaussian width, retaining the switching and time ordering.

We work with a massless, conformally coupled scalar field in Minkowski, global anti-de Sitter (AdS), and de Sitter (dS) spacetimes, whose two-point functions are explicit for every $\mathcal D\ge3$. The flat-space result gives a check on the curved-space calculations. AdS adds boundary and redshift effects. The Bunch--Davies state is thermal in the dS static patch. The Ba\~nados--Teitelboim--Zanelli (BTZ) black hole is locally AdS, but its quotient images change the detector response \cite{HodgkinsonLouko2012}. This tests global geometry at fixed local curvature. Constant scalar curvature alone would not give the kernels used here.

For static pairs with equal redshifts, local responses, Gaussian widths and gaps, the mean-time integral factors the gap dependence out of $X$. In dS, Kubo--Martin--Schwinger (KMS) detailed balance fixes the gap maximizing $|X|/P$, with $P=P_A=P_B$. The existence test needs only $X(0)$ and the local response at that gap. The leading entangling interval is bounded; a nonempty vacuum interval in Minkowski or global AdS has no upper endpoint.

We find a logarithmic law for the BTZ response growth seen numerically near the singularity \cite{PhysRevD.110.025002}. A uniform image-sum estimate gives the coefficient. It is positive for nonnegative, nondecreasing onsets with nonzero terminal amplitude. If the onset extends smoothly by zero before release, the finite-time glitches disappear, but the logarithm remains. The readout is still sharp.

We find that our analytic results provide new insights into two RQI phenomena. First, our 
thermal result answers a question regarding whether tuning the detectors' energy gap can produce entanglement at all. Given purely numerical methods, one might ask if, for a given fixed detector arrangement and switching width, can {\it any} choice of energy gap produce entanglement, or is searching over gaps futile? In Part E of Section V  we answer this question: there is a   distinguished energy gap $\Omega_* = \beta/4\sigma^2$.
The possibility of entanglement hinges on a comparison criterion involving only $X(0)$ and $P(\Omega_*)$. If that comparison fails, tuning the energy gap cannot yield entanglement. This is a decision criterion about the existence of entanglement in the final detector state, not merely an expression for a response curve.

Further new insights concern the BTZ black hole, where we obtain  answers as to what the infalling detector response curve means. Determining whether the growth in the transition rate near the singularity requires a sudden switch-on can be difficult using purely numerical methods. However, our analytic results show that the transition rate still diverges logarithmically for the smooth onsets described above, even though the finite-time glitches disappear. The readout remains sharp. We thus learn  which feature of the curve survives the change of switching onset (from sharp to smooth). Our analytic approach traces this divergence to the accumulation of quotient-image contributions; the covering-AdS contribution remains finite.

Earlier analytic and spectral treatments of AdS and dS \cite{henderson2019entangling,NgMannMartinMartinez2018,Lin_2024} provide the starting point. We keep the finite-width dependence and each coefficient's boundary prescription. The formulas complement single-detector magic-resource calculations \cite{Yang2025analytic,Zhang:2026jll} and give benchmarks for analogue detectors and field-correlation studies \cite{Gooding:2020scc,Settembrini:2021cye,Lindel:2023rfi,Gooding:2023xxl,Gooding:2025tfp,Gooding:2026nkm}.  After the conventions in Sec.~\ref{sec:udw_and_rho}, Secs.~\ref{sec:Minkowski}--\ref{sec:BTZ} treat Minkowski, AdS, dS, and BTZ spacetimes in order. Supporting derivations and the integral lookup table are in the appendices.

\section{UDW detectors and reduced density matrix elements}
\label{sec:udw_and_rho}

Throughout, $\mathcal D=d+1=2p+2$ is spacetime dimension, $d\ge2$ is spatial dimension, and $p=(d-1)/2>0$. We use signature $(-,+,\ldots,+)$ and $\hbar=c=1$. The field operator $\phi$ obeys $(\Box-\xi_{\mathcal D}\mathcal R)\phi=0$, where $\mathcal R$ is scalar curvature and $\xi_{\mathcal D}=(\mathcal D-2)/[4(\mathcal D-1)]$.

Each detector has energy eigenstates $\ket{0_D},\ket{1_D}$ and monopole operator
\begin{equation}
\mu_D(\tau_D)=\sigma_D^+e^{i\Omega_D\tau_D}
+\sigma_D^-e^{-i\Omega_D\tau_D},\quad
\sigma_D^+=\ket{1_D}\!\bra{0_D},
\end{equation}
with $\sigma_D^-=(\sigma_D^+)^\dagger$. Its interaction Hamiltonian in proper time is
\begin{equation}\label{eq:UDW_Hamiltonian}
H_D(\tau_D)=\lambda\chi_D(\tau_D)\mu_D(\tau_D)
\otimes\phi(x_D(\tau_D)).
\end{equation}
Here, $\lambda$ is the coupling strength, $\chi_D(\tau_D)$ is the switching function, $\phi(x_D(\tau_D))$ is the field operator evaluated along the detector worldline $x_D(\tau_D)$, and $\tau_D$ is the detector proper time. The initial state is $\ket{0_A0_B}\!\bra{0_A0_B}\otimes\rho_\phi$. The field state is specified in each application. The dS static-patch and BTZ exterior restrictions are thermal with respect to their Killing flows, not empty static vacua. We order the Dyson expansion in $t$. On static worldlines,
\begin{equation}\label{eq:tau_eta_t}
\tau_D=\eta_D t,\quad \eta_D=\sqrt{-g_{tt}(\mathbf x_D)},
\end{equation}
and
\begin{equation}\label{eq:U_time_ordered}
U=\mathcal T\exp\!\left[-i\int dt\sum_{D=A,B}
\frac{d\tau_D}{dt}H_D(\tau_D)\right].
\end{equation}
Tracing out the field gives, in the basis $(\ket{00},\ket{01},\ket{10},\ket{11})$,
\begin{equation}\label{eq:rho_X_state}
\rho_{AB}=\begin{pmatrix}
1-P_A-P_B&0&0&X\\
0&P_B&C&0\\
0&C^*&P_A&0\\
X^*&0&0&0
\end{pmatrix}+\mathcal O(\lambda^4).
\end{equation}
In particular, $X=\rho_{14}$ is the conjugate of the double-excitation amplitude often denoted $\mathcal M$, and $C=\rho_{23}$ is the single-excitation coherence. Define
\begin{equation}\label{eq:Wightman_def}
W(x,x')=\operatorname{Tr}[\rho_\phi\phi(x)\phi(x')].
\end{equation}
Writing $W(\tau_A,\tau_B)$ for its worldline pullback, the leading terms are
\begin{multline}\label{eq:PD_general}
P_D=\lambda^2\!\int d\tau_D\,d\tau_D'\,
\chi_D(\tau_D)\chi_D(\tau_D')e^{-i\Omega_D(\tau_D-\tau_D')}\\
\times W(x_D(\tau_D),x_D(\tau_D')),
\end{multline}
\begin{multline}\label{eq:C_general}
C=\lambda^2\!\int d\tau_A\,d\tau_B\,
\chi_A(\tau_A)\chi_B(\tau_B)e^{-i(\Omega_A\tau_A-\Omega_B\tau_B)}\\
\times W(\tau_A,\tau_B),
\end{multline}
and
\begin{multline}\label{eq:X_general}
X=-\lambda^2\!\int d\tau_A\,d\tau_B\,
\chi_A(\tau_A)\chi_B(\tau_B)e^{-i(\Omega_A\tau_A+\Omega_B\tau_B)}\\
\times\bigl[\theta(t_B-t_A)W(\tau_A,\tau_B)
+\theta(t_A-t_B)W(\tau_B,\tau_A)\bigr].
\end{multline}
Here $t_D=t_D(\tau_D)$ and $\theta$ is the Heaviside function. The earlier field operator is first in each contribution to $X$. Thus a positive time-difference reduction uses the upper time boundary, $W(-s)$, whereas the full-line $C$ integral uses the usual lower boundary. Complex switching coefficients at nonzero lag must not be conjugated when fixing this prescription.

For the static calculations we take equal response frequencies $\Omega_A=\Omega_B=\Omega$ and a common proper switching width $\sigma>0$. The coupling has dimension $[\lambda]=L^{p-1}$, and $g=\lambda\sigma^{1-p}$ is dimensionless. We use the switching functions~\cite{henderson2019entangling}
\begin{equation}\label{eq:Gaussian_switching}
\begin{aligned}
\chi_A(\tau_A)&=e^{-(\tau_A+\eta_A t_0/2)^2/(2\sigma^2)},\\
\chi_B(\tau_B)&=e^{-(\tau_B-\eta_B t_0/2)^2/(2\sigma^2)},
\end{aligned}
\end{equation}
Positive $t_0$ means that detector $A$ peaks before $B$ in Killing time, with $\tau_D=\eta_Dt$ as in Eq.~\eqref{eq:tau_eta_t}. Ground-state excitation has $\Omega>0$; negative signed frequencies in response plots describe the de-excitation continuation. For $X$ we assume positive spatial separation. At coincidence, time ordering meets the ultraviolet singularity and smooth switching alone does not define the pointlike cross term, already in $\mathcal D=3$.

The leading concurrence is
\begin{equation}\label{eq:concurrence}
\mathcal C[\rho_{AB}]=2\max\{0,|X|-\sqrt{P_AP_B}\}
+\mathcal O(\lambda^4).
\end{equation}
This follows from the perturbative physical state, including its higher-order completion; the displayed second-order matrix is not asserted to be exactly positive by itself. The coherence $C$ affects detector correlations but does not enter this leading concurrence. Gaussian tails do not enforce strictly spacelike separated interactions, so the plotted entanglement need not isolate a signaling-free harvesting contribution. Plots divided by $g^2$ or $\lambda^2$ show perturbative coefficients, not a claim of validity at unit physical coupling.

\section{Minkowski vacuum: the finite-width baseline}
\label{sec:Minkowski}

We first calculate the response of static detectors in the Minkowski vacuum. There is no curvature, gravitational redshift, or reflecting boundary. Finite-switching harvesting in flat spacetime was studied in Ref.~\cite{PozasKerstjens:2015}. Here we give the coefficients in arbitrary dimension, with their causal boundary prescriptions. The four-dimensional formulas check the normalization and phases. The vacuum spectrum will also provide the contrast with dS in Sec.~\ref{subsec:thermal_gap}.

\subsection{Wightman function}
\label{subsec:Mink_Wightman}

For the massless scalar Minkowski vacuum in $\mathcal D=d+1=2p+2\ge3$ dimensions,
\begin{align}
 W(x,x')&=\mathcal A_p
 \bigl[|\mathbf x-\mathbf x'|^2-(t-t'-i0)^2\bigr]^{-p},
 \\[-2pt]
 \mathcal A_p&=\frac{\Gamma(p)}{4\pi^{p+1}}.
 \label{eq:Mink_Wightman}
\end{align}
Here $i0$ denotes the boundary value from $\epsilon>0$, with the power continued from spacelike separation~\cite{Smith.2017,LoukoToussaint2016}. For static detectors the spatial separation is $L$. We use the switchings of Sec.~\ref{sec:udw_and_rho}, for which $\eta_A=\eta_B=1$, and write $q=\sigma\Omega$. The $X$ formulas below require $L>0$; coincident $P$ and $C$ remain well defined.

\subsection{Excitation probability}
\label{subsec:Mink_probability}

The positive-frequency identity
\begin{align}
 e^{-i\pi p}(s-i\epsilon)^{-2p}
 &=\frac1{\Gamma(2p)}\int_0^\infty
 k^{2p-1}e^{-iks-\epsilon k}\,\mathrm dk
 \label{eq:Mink_P_spectral_identity}
\end{align}
licenses the Gaussian transform before taking $\epsilon\to0^+$. It gives
\begin{align}
 P=\frac{\lambda^2\Gamma(p)\sigma^2}{2\pi^p\Gamma(2p)}
 \int_0^\infty k^{2p-1}e^{-\sigma^2(k+\Omega)^2}\,\mathrm dk.
 \label{eq:Mink_P_positive}
\end{align}
This integral is positive and converges for every real $\Omega$. The Gaussian averages the spectrum over a width of order $\sigma^{-1}$. Finite-width excitation is therefore possible even though the stationary excitation rate vanishes for an inertial detector in vacuum.

Expanding $e^{-2\sigma^2\Omega k}$ into its even and odd parts and applying Kummer's transformation yields
\begin{align}
 P={}&\frac{\lambda^2\sigma^{2-2p}}{2^{2p+1}\pi^{p-1/2}}
 \Bigg[\frac{\Gamma(p)}{\Gamma(p+\tfrac12)}
 {}_1F_1\!\left(\tfrac12-p;\tfrac12;-q^2\right)\nonumber\\
 &\hspace{4em}-2q\,{}_1F_1\!\left(1-p;\tfrac32;-q^2\right)\Bigg].
 \label{eq:excitationMinkclosedform}
\end{align}
Equivalently, in terms of the parabolic-cylinder function $D_\nu$,
\begin{align}
 P=\frac{\lambda^2\Gamma(p)\sigma^{2-2p}}{2^{p+1}\pi^p}
 e^{-q^2/2}D_{-2p}(\sqrt2q).
 \label{eq:Mink_P_cylinder}
\end{align}
Both expressions include $q=0$ and negative signed response frequencies. For $p=1$ they reduce to~\cite{Smith.2017}
\begin{align}
 P=\frac{\lambda^2}{4\pi}
 \left[e^{-q^2}-\sqrt\pi q\,\erfc(q)\right].
 \label{eq:Mink_P_fourdimensional}
\end{align}
A Tricomi representation obtained using
$D_{-2p}(\sqrt2q)=2^{-p}q e^{-q^2/2}U(p+\tfrac12,\tfrac32,q^2)$ requires $q>0$; this identity must not be extended to negative $q$.

\subsection{Correlation term: \texorpdfstring{$X$}{X}}
\label{subsec:Mink_X}

At $t_0=0$, integration over the mean time in Eq.~\eqref{eq:X_general} gives
\begin{align}
 X=-2\lambda^2\mathcal A_p\sqrt\pi\sigma e^{-q^2}
 \int_0^\infty\frac{e^{-s^2/(4\sigma^2)}\,\mathrm ds}
 {[L^2-(s+i0)^2]^p}.
 \label{eq:Mink_X_halfline}
\end{align}
The upper boundary follows from $X=\rho_{14}$. To evaluate the integral, first substitute $y=s^2/z$ in
$\int_0^\infty e^{-a s^2}(s^2+z)^{-p}\,\mathrm ds$ for $a>0$, $z>0$. The Euler integral and transformation of $U$ give
$\tfrac12\sqrt\pi a^{p-1/2}U(p,p+\tfrac12,az)$.
Continue to $z=-L^2+i0$ and multiply by $e^{i\pi p}$. Thus
\begin{align}
 X=-\frac{\lambda^2\Gamma(p)e^{i\pi p}\sigma^{2-2p}}{2^{2p+1}\pi^p}
 e^{-q^2}\times U\!\left(p,p+\tfrac12,-\frac{L^2}{4\sigma^2}+i0\right).
 \label{eq:Mink_X_closed}
\end{align}
The lip of the cut is part of this formula. In $\mathcal D=4$,
\begin{align}
 X=-\frac{\lambda^2\sigma}{4\sqrt\pi L}
 e^{-q^2-L^2/(4\sigma^2)}
 \left[\erfi\!\left(\frac L{2\sigma}\right)+i\right].
 \label{eq:Mink_X_fourdimensional}
\end{align}
For nonzero lag, replace the Gaussian in the integral of Eq.~\eqref{eq:Mink_X_halfline} by
$e^{-s^2/(4\sigma^2)-t_0^2/(4\sigma^2)}\cosh[st_0/(2\sigma^2)]$.
The coincident limit of Eq.~\eqref{eq:Mink_X_closed} is divergent, already logarithmically at $p=1/2$.

\subsection{Correlation term: \texorpdfstring{$C$}{C}}
\label{subsec:Mink_C}

Distinguish the bare kernel integral
\begin{align}
 C_{\mathrm M}(p;L;a,b)
 &=\int_{-\infty}^\infty
 \frac{e^{-as^2-ibs}\,\mathrm ds}{[L^2-(s-i0)^2]^p},
 \label{eq:Mink_C_integral}
\end{align}
from the physical single-excitation coherence. For our switchings,
\begin{align}
 C&=\lambda^2\mathcal A_p\sqrt\pi\sigma
 e^{-t_0^2/(4\sigma^2)}C_{\mathrm M}(p;L;a,b),
 \label{eq:Mink_C_physical}\\
 a&=\frac1{4\sigma^2},\qquad
 b=\Omega-\frac{it_0}{2\sigma^2}.
 \label{eq:Mink_C_parameters}
\end{align}
The complex $b$ encodes the lag. Appendix~\ref{app:Mink_C} gives a convergent spectral master integral, the exact $p=1/2,1$ base cases, and the dimension-raising rule. In particular, for $\mathcal D=4$ and $t_0=0$,
\begin{align}
 C=\frac{\lambda^2\sigma}{4\sqrt\pi L}\Im\!\left[
 e^{-L^2/(4\sigma^2)-i\Omega L}
 \erfc\!\left(q-\frac{iL}{2\sigma}\right)\right].
 \label{eq:Mink_C_fourdimensional}
\end{align}
At coincident identical simultaneous switchings, the spectral representation gives $C=P$.

\subsection{Vacuum benchmark and numerical checks}
\label{subsec:Mink_benchmark}

At fixed separation and lag, $X$ carries the factor $e^{-\sigma^2\Omega^2}$. Dividing the local response in Eq.~\eqref{eq:Mink_P_positive} by this factor leaves a strictly decreasing function of gap that tends to zero. Thus, if $X(0)\ne0$, the leading entanglement criterion holds at all sufficiently large gaps. The concurrence nevertheless tends to zero. Section~\ref{subsec:thermal_gap} proves this vacuum result and its thermal counterpart; neither controls the full detector state uniformly at large gap.

\begin{figure*}[t]
 \centering
 \includegraphics[width=0.78\textwidth]{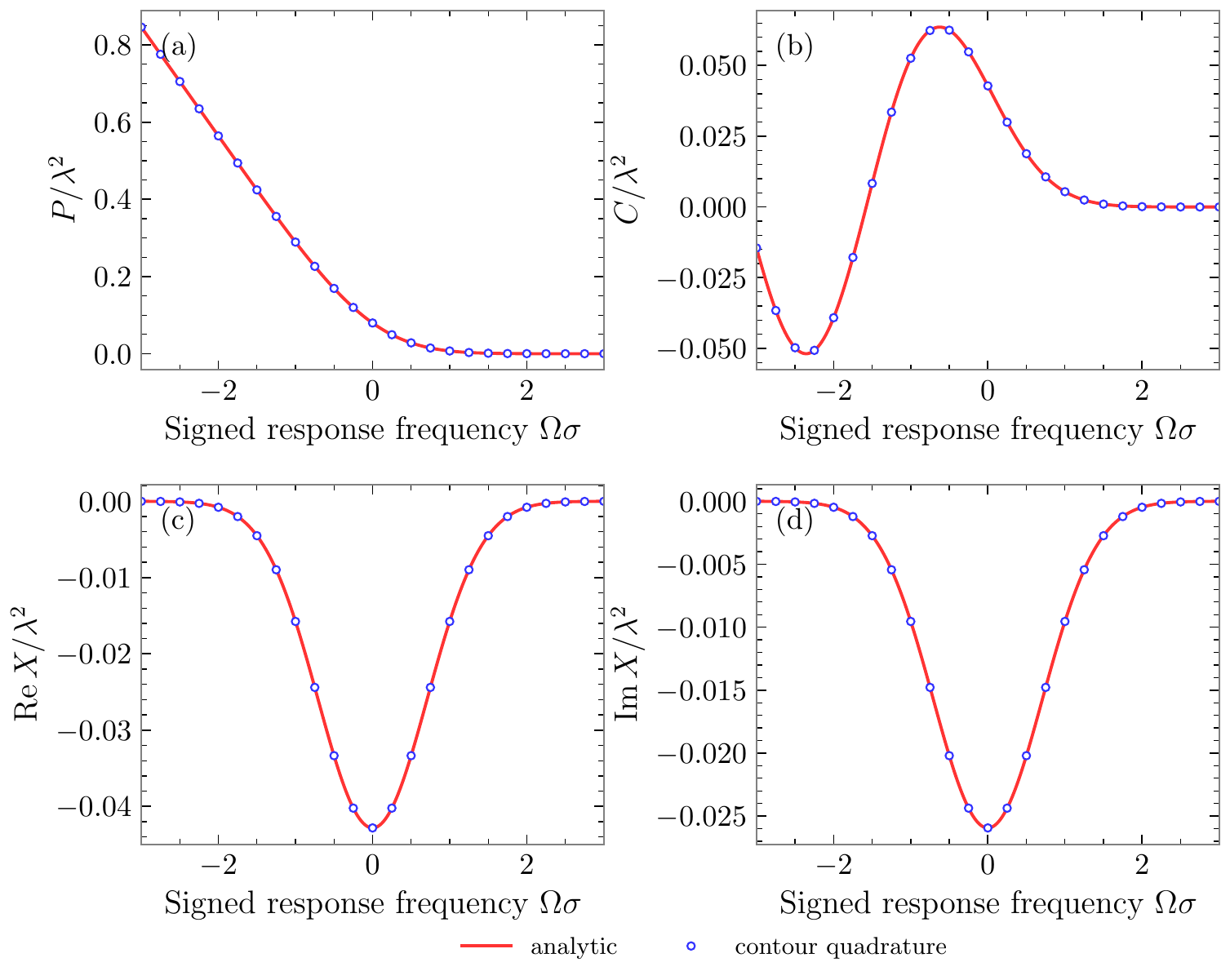}
 \caption{Minkowski $\mathcal{D}=4$ response coefficients for $L=2\sigma$ and $t_0=0$: (a) $P/\lambda^2$, (b) $C/\lambda^2$, (c) $\Re X/\lambda^2$, and (d) $\Im X/\lambda^2$. Analytic curves are compared with 25 independent contour-quadrature samples over $-3\le\Omega\sigma\le3$. Negative signed frequencies describe the de-excitation response; positive frequencies describe excitation from the ground state. The sign of $\Im X$ corresponds to $X=\rho_{14}$.}
 \label{fig:Minkowski_probability_correlations_check}
\end{figure*}

Fig. \ref{fig:Minkowski_probability_correlations_check} tests the signed real and imaginary parts as well as $P$ and $C$.
The independent numerical integrals use the contour deformations in Appendix~\ref{app:Mink_C}; no finite regulator is fitted. For the plotted samples the largest absolute difference is $5.6\times10^{-16}$, and changing the contour displacement from $\sigma$ to $\sigma/2$ changes the checked values by less than $3.4\times10^{-16}$. These double-precision checks test the normalization and branch choice, without implying uniform error bounds in dimension or separation.

\section{Global AdS: redshift and boundary conditions}
\label{sec:AdS}

Boundary conditions matter in global AdS, as does the redshift between static detectors. Studies in AdS$_3$ and AdS$_4$ used integral reductions and mode decompositions to examine both effects \cite{henderson2019entangling,NgMannMartinMartinez2018}. We construct the static two-detector coefficients in every $\mathcal D\ge3$. The  difficulty is $X$: time ordering leaves an algebraic mode tail, unlike the Gaussian tails of $P$ and $C$. Subtracting this tail gives convergent expressions at separated positions.

\subsection{Wightman function}
\label{subsec:AdS_Wightman}

For a massless conformally coupled scalar in the global AdS vacuum~\cite{henderson2019entangling,Jennings:2010vk},
\begin{align}
 W(x,x')&=\frac{A_d}{\ell^{2p}}
 \left[\Sigma^{-p}-\zeta(\Sigma+2)^{-p}\right],\,
 A_d=\frac{\Gamma(p)}{2^{p+2}\pi^{p+1}}.
\end{align}
Here $\Sigma=(X-X')^2_{\mathbb R^{2,d}}/(2\ell^2)$ is the dimensionless embedding chordal invariant, not the intrinsic geodesic distance squared. $\ell$ is the AdS radius. The choices $\zeta=-1,0,1$ denote Neumann, transparent and Dirichlet conditions, respectively; general Robin conditions are not included. For static positions with angular separation $\theta$,
\begin{multline}
 \Sigma=-1+\frac{\gamma_A\gamma_B}{\ell^2}
 \cos\!\left(\frac{t_A-t_B}{\ell}\right)
 -\frac{R_AR_B}{\ell^2}\cos\theta,\\
 \gamma_D=\sqrt{\ell^2+R_D^2},\qquad
 t_D=\frac{\ell\tau_D}{\gamma_D}.
\end{multline}
The fractional powers require a global prescription. The Gegenbauer generating function
\begin{equation}\label{eq:generating_function}
 (1-2xy+y^2)^{-p}=\sum_{n=0}^{\infty}C_n^p(x)y^n,
 \qquad |y|<1
\end{equation}
defines the positive-frequency branch on the universal cover by
\begin{equation}\label{eq:AdS_branch}
 F_p(z,x):=2^p\sum_{n=0}^{\infty}C_n^p(x)e^{-i(n+p)z},
 \qquad \Im z<0.
\end{equation}
This is the analytic meaning of $(\cos z-x)^{-p}$ for $-1\le x\le1$. It obeys $F_p(z+2\pi,x)=e^{-2\pi ip}F_p(z,x)$; taking a fresh principal power of the cosine would lose this phase. The static pullback is therefore
\begin{multline}\label{eq:two_static_AdS}
 W(\tau_A,\tau_B)=\frac{A_d}{(\gamma_A\gamma_B)^p}
 \bigl[F_p(\varphi-i0,\alpha_+)\\
 -\zeta F_p(\varphi-i0,-\alpha_-)\bigr],
\end{multline}
where $\varphi=\tau_A/\gamma_A-\tau_B/\gamma_B$ and
$\alpha_\pm=(\ell^2\pm R_AR_B\cos\theta)/(\gamma_A\gamma_B)$.
For one detector, $\alpha_+=1$ and $\alpha_-=\mu=(\ell^2-R^2)/\gamma^2$.

\subsection{Excitation probability}
\label{subsec:AdS_probability}

Stationarity makes the probability independent of the center of its Gaussian switching. Integrating the mean proper time and putting $s=(\tau-\tau')/\gamma$ gives
\begin{multline}\label{eq:AdS_P_after_pullback}
 P=\sqrt\pi\sigma\lambda^2A_d\gamma^{1-2p}
 \int_{\mathbb R}ds\,e^{-\gamma^2s^2/(4\sigma^2)-i\Omega\gamma s}\\
 \times\left[F_p(s-i0,1)-\zeta F_p(s-i0,-\mu)\right].
\end{multline}
In particular, $F_p(s-i\epsilon,1)=2^{-p}[i\sin((s-i\epsilon)/2)]^{-2p}$ on the prescribed branch. Equation~\eqref{eq:AdS_branch} reduces both terms to Gaussian Fourier transforms, evaluated at $\epsilon>0$ before taking $\epsilon\downarrow0$. Writing $\omega_n=n+p$, $P=P_0-\zeta P_b$, and $K_P=\lambda^2\Gamma(p)\sigma^2/(2\pi^p\gamma^{2p})$, one obtains
\begin{align}
 P_0&=K_P\sum_{n=0}^{\infty}\frac{(2p)_n}{n!}
 e^{-\sigma^2(\Omega+\omega_n/\gamma)^2},
 \label{eq:AdS_P0_def}\\
 P_b&=K_P\sum_{n=0}^{\infty}C_n^p(-\mu)
 e^{-\sigma^2(\Omega+\omega_n/\gamma)^2}.
 \label{eq:AdS_Pb_def}
\end{align}
Here $(u)_n=\Gamma(u+n)/\Gamma(u)$. Since $C_n^p(1)=(2p)_n/n!$, the total probability is
\begin{align}\label{eq:AdSexcitation}
    P=\frac{\lambda^2\Gamma(p)\sigma^2}{2\pi^p\gamma^{2p}}
 \sum_{n=0}^{\infty}\left[C_n^p(1)-\zeta C_n^p(-\mu)\right]
 \times e^{-\sigma^2(\Omega+\omega_n/\gamma)^2}.
\end{align}
Gaussian damping makes this series absolutely convergent for every finite static radius and real signed frequency $\Omega$.

Taking $\ell\to\infty$ in Eq.~\eqref{eq:AdSexcitation}, at fixed detector position and switching parameters, recovers the Minkowski spectral integral~\eqref{eq:Mink_P_positive}.

\subsection{Correlation term: \texorpdfstring{$X$}{X}}
\label{subsec:AdS_X}

The convention $X=\rho_{14}$ in Eq.~\eqref{eq:X_general} places the earlier field operator first. After integrating the mean coordinate time and pairing the two time orderings, the kernel is consequently $W(-s)$, with $s=|t_A-t_B|/\ell$. With $f(s)=e^{-a_2s^2}\cos(a_3s)$, define
\begin{equation}\label{eq:AdS_X_integral}
 X_0=a_1\int_0^\infty ds\,f(s)F_p(-s-i0,\alpha_+).
\end{equation}
The boundary contribution $X_b$ follows by $\alpha_+\to-\alpha_-$, and $X=X_0-\zeta X_b$. The coefficients $a_1,a_2,a_3$ are given explicitly in Appendix~\ref{app:coefficients}; $a_1$ includes $\lambda^2$ and the generally complex lag factor. The reflected kernel in~\eqref{eq:AdS_X_integral} is the upper boundary $(\cos(s+i0)-\alpha_+)^{-p}$, so its modes are $e^{+i\omega_ns}$. Writing $J_n^+=\int_0^\infty f(s)e^{i\omega_ns}ds$, their half-Gaussian transforms are
\begin{align}\label{eq:AdS_Jn_closed}
 J_n^+&=\frac{\sqrt\pi}{4\sqrt{a_2}}\sum_{\xi=\pm1}
 e^{-z_{n\xi}^2}\left[1+i\erfi(z_{n\xi})\right],\\
 z_{n\xi}&=\frac{\omega_n+\xi a_3}{2\sqrt{a_2}}.\notag
\end{align}
For numerical stability the bracketed product can be evaluated as $e^{-z^2}\erfc(-iz)$. At fixed $-1<\alpha_+<1$, the raw series $\sum_n C_n^p(\alpha_+)J_n^+$ is absolutely convergent for $p<1$, conditionally convergent for $1\le p<2$, and generally divergent for $p\ge2$. Repeated integration by parts supplies the subtraction
\begin{equation}
 J_n^{+,\mathrm{asy}}(N)=\sum_{j=0}^{N}
 \frac{f^{(2j)}(0)}{(-i\omega_n)^{2j+1}},\quad N=\lceil p\rceil,
\end{equation}
with remainder $O(n^{-2N-3})$. Thus
\begin{multline}\label{eq:AdS_X0_final}
 X_0=2^pa_1\sum_{n=0}^{\infty}C_n^p(\alpha_+)
 \left[J_n^+-J_n^{+,\mathrm{asy}}(N)\right]\\
 +2^pa_1\sum_{j=0}^{N}f^{(2j)}(0)e^{i\pi(2j+1)/2}
 S_{2j+1}(\alpha_+).
\end{multline}
Appendix~\ref{app:AdS_sums} gives the derivatives and convergent integral representations of the restored Abel sums $S_{2j+1}$. The subtracted series is absolutely convergent for all $d\ge2$ and kernel arguments in $[-1,1)$. Spatial coincidence, where $\alpha_+=1$, is excluded for pointlike $X$: its half-line ultraviolet singularity requires smearing or a specified renormalized extension. This restriction does not apply to $P$ or $C$. 

Equation~\eqref{eq:AdS_X0_final} defines a convergent mode representation where the raw series fails. Restoring the Abel moments keeps the boundary distribution unchanged. This does not renormalize the excluded coincident pointlike $X$.

\subsection{Correlation term: \texorpdfstring{$C$}{C}}
\label{subsec:AdS_C}

Unlike $X$, Eq.~\eqref{eq:C_general} factorizes into two Gaussian Fourier transforms after insertion of~\eqref{eq:AdS_branch}. Each mode shifts the two proper frequencies to $\Omega+\omega_n/\gamma_D$; their switching centers supply the lag phase. With $a=\gamma_A$, $b=\gamma_B$, $\delta=t_0/\ell$, and
$G_n=C_n^p(\alpha_+)-\zeta C_n^p(-\alpha_-)$, the result is
\begin{multline}\label{eq:AdS_C_final}
 C=\frac{\lambda^2\Gamma(p)\sigma^2}{2\pi^p(ab)^p}
 \sum_{n=0}^{\infty}G_n
 e^{i\delta[\omega_n+\Omega(a+b)/2]}\\
 \times\exp\!\left[-\frac{\sigma^2}{2}
 \left\{\left(\Omega+\frac{\omega_n}{a}\right)^2+
 \left(\Omega+\frac{\omega_n}{b}\right)^2\right\}\right].
\end{multline}
The series is absolutely convergent, including at coincidence. For the same worldline and $t_0=0$, it reduces term by term to~\eqref{eq:AdSexcitation}. Positivity of the field two-point function also implies $|C|^2\le P_AP_B$.

\subsection{Comparison with numerical results}
\label{subsec:AdS_comparison}

\begin{figure}[!tbp]
 \centering
 \includegraphics[width=0.95\columnwidth]{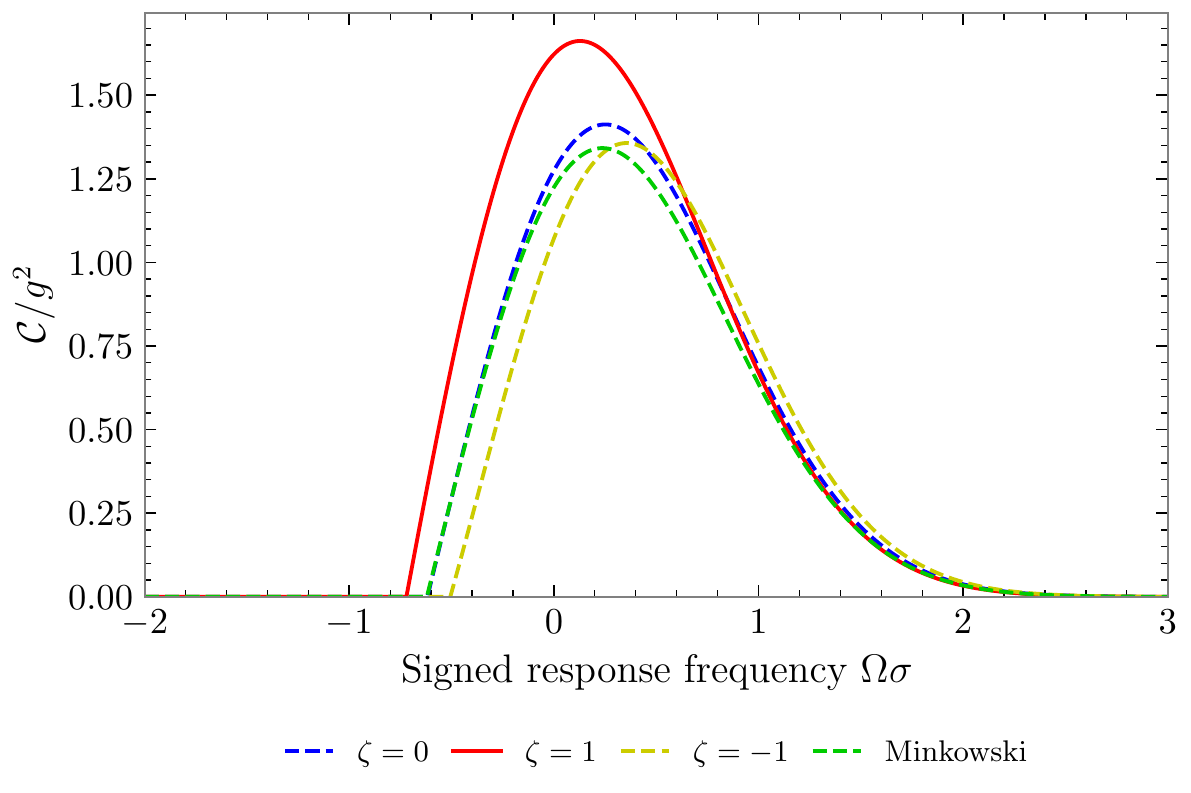}
 \caption{Concurrence normalized by $g^2=\lambda^2\sigma$ in $\mathrm{AdS}_3$, with $\ell=\sigma$, $R_A=0$, proper separation $L=0.1\sigma$, $R_B=\ell\sinh(L/\ell)$, and $t_0=0$. The three boundary conditions and the Minkowski comparator are shown; cf.\ Fig.~5 of Ref.~\cite{henderson2019entangling}. Negative $\Omega$ denotes the signed-response branch discussed in Sec.~\ref{sec:udw_and_rho}.}
 \label{fig:AdS_Concurrence_Energy}
\end{figure}

Figure~\ref{fig:AdS_Concurrence_Energy} uses the zero-lag parameters of Fig.~5 in Ref.~\cite{henderson2019entangling}. It reproduces the boundary-dependent concurrence and compares it with the Minkowski result from Eqs.~\eqref{eq:Mink_P_positive} and~\eqref{eq:Mink_X_closed}. The boundary conditions change the peak and gap dependence; all displayed curves decay toward zero at large positive gap.

\begin{figure}[!tbp]
 \centering
 \includegraphics[width=0.95\columnwidth]{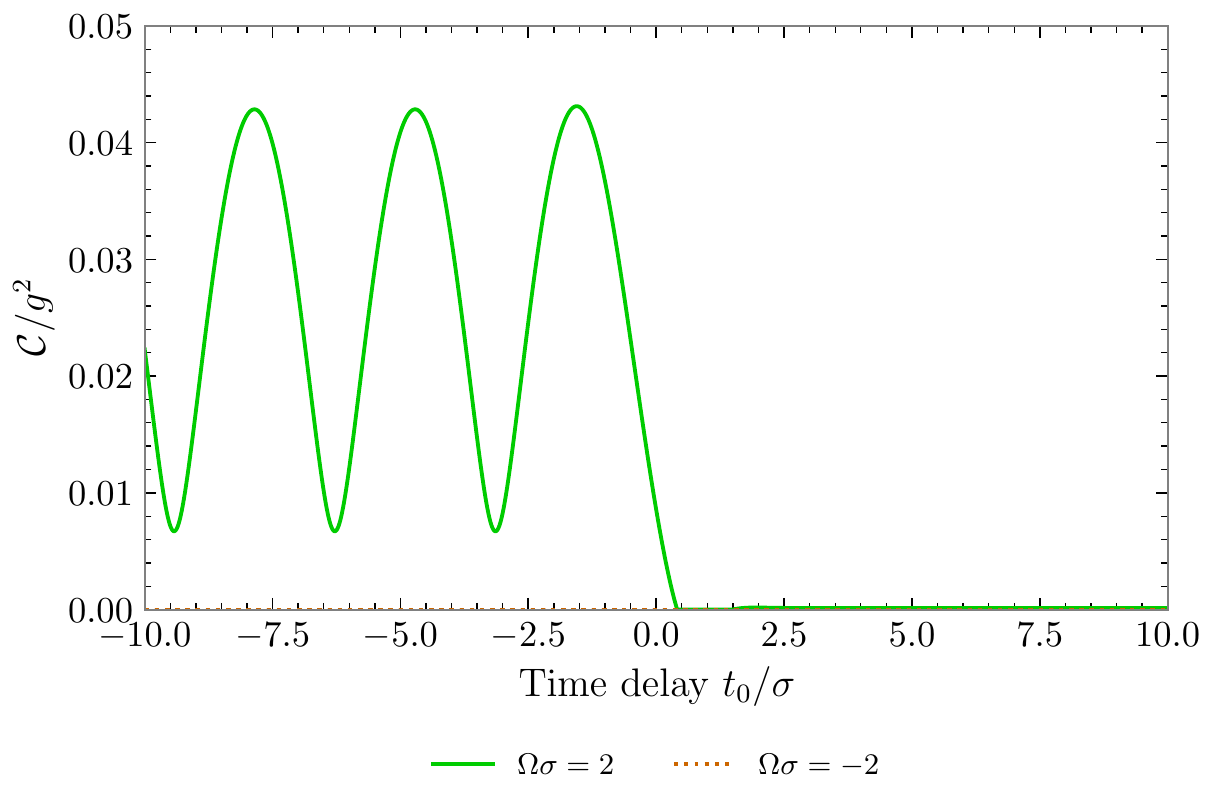}
 \caption{Concurrence versus coordinate-time delay, with $\ell=\sigma$, $R_A=0$, $L=5\sigma/2$, $R_B=\ell\sinh(L/\ell)$, and Neumann conditions. Positive $t_0$ means that $A$ peaks first. The $\Omega\sigma=2$ curve uses the corrected earlier-first Wightman ordering; the $\Omega\sigma=-2$ curve vanishes throughout this range. The lag orientation relative to Fig.~11 of Ref.~\cite{henderson2019entangling} is explained in the text.}
 \label{fig:AdS_Concurrence_Time_Delay}
\end{figure}

The analytic calculation uses 501 modes for this figure and 301 for Fig.~\ref{fig:AdS_Concurrence_Time_Delay}. For the latter parameters, independent branch-tracked quadrature and the subtracted complex $X$ agree within $4\times10^{-15}$ in $X/g^2$ at $t_0/\sigma=-5,-2,0,2,5$ in double precision.

With our switchings and $X=\rho_{14}$, the ordered kernel is $W(-s)$. Figure~\ref{fig:AdS_Concurrence_Time_Delay} is therefore reflected in $t_0$ relative to Fig.~11 of Ref.~\cite{henderson2019entangling}. That reference's Eqs.~(2.6) and (B.16) place the earlier operator first, but Eq.~(B.17) uses the opposite boundary value. Here the lag asymmetry comes from unequal redshifts. These configurations do not satisfy the assumptions of the gap criterion in Sec.~\ref{subsec:thermal_gap}.


\section{Static-patch de Sitter: thermal constraints}
\label{sec:dS}

The Bunch--Davies state is thermal under static Killing evolution, but a single detector's thermal response need not determine two-detector correlations. Ver Steeg and Menicucci compared inertial pairs in an expanding dS patch with pairs in a thermal Minkowski environment \cite{VerSteegMenicucci2009}; their trajectories differ from ours. Lin and Mondal treated a broader family of dS vacua at zero and antipodal separations, using spectral methods and a large-measuring-time saddle-point approximation \cite{Lin_2024}.

We use the Bunch--Davies state and distinct static worldlines within one patch. The response formulas keep the exact Gaussian-width dependence, with normalized dimension raising to all $\mathcal D\ge3$. For a fixed equal-redshift pair, we then ask whether any gap gives leading-order entanglement. Spectral positivity and KMS detailed balance give an exact test.

\subsection{Wightman function}
\label{subsec:dS_Wightman}

Let $\alpha>0$ be the dS curvature radius, so the expansion rate is $H=\alpha^{-1}$. In the static patch, $0\le r<\alpha$ and $\gamma=\sqrt{\alpha^2-r^2}$. For a massless conformally coupled scalar in the Bunch--Davies state, the two-point function is \cite{Lin_2024,BezerraSaharian2009}
\begin{align}
 W(x,x')&=\frac{\Gamma(p)\alpha^{-2p}}{(4\pi)^{p+1}}c(x,x')^{-p},
 \qquad c=\frac{1-Z}{2},\nonumber\\
 Z&=\frac{\gamma\gamma'\cosh(\Delta t/\alpha)+rr'\cos\Theta}{\alpha^2}.
\end{align}
Here $Z$ is the dimensionless embedding invariant and $\Theta$ is the angular separation. The causal boundary value gives
\begin{equation}
 W(x,x')=\frac{\Gamma(p)(\gamma\gamma')^{-p}}{2^{p+2}\pi^{p+1}}
 K_p^-(\Delta t/\alpha;\kappa),
 \label{eq:dS_W_static}
\end{equation}
where
\begin{align}
 K_p^\pm(s;\kappa)&:=[\kappa-\cosh(s\pm i0)]^{-p},\nonumber\\
 \kappa&=\frac{\alpha^2-rr'\cos\Theta}{\gamma\gamma'}.
 \label{eq:dS_kernel}
\end{align}
The powers denote analytic boundary values. The static-patch restriction of this state is thermal with respect to Killing time; it is not the empty static vacuum.

\subsection{Excitation probability}
\label{subsec:dS_probability}

At radius $R$, proper time is $\tau=\gamma t/\alpha$, with $\gamma=\sqrt{\alpha^2-R^2}$. After the Gaussian mean-time integration, introduce a real auxiliary shift $x$:
\begin{align}
 P(p;x)&=\mathcal N_p\int_{\mathbb R}e^{-As^2-iBs}
 \csch^{2p}\!\left(\frac{s+x-i0}{2}\right)\,ds,
 \label{eq:dS_P_integral}\\
 A&=\frac{\gamma^2}{4\sigma^2},\qquad B=\gamma\Omega,\nonumber\\
 \mathcal N_p&=\frac{e^{-i\pi p}\sqrt\pi\sigma\lambda^2\Gamma(p)}{(4\pi)^{p+1}}
 \gamma^{1-2p},\qquad P=P(p;0).\nonumber
\end{align}
Writing the hyperbolic factor as $k_p(s+x)$, direct differentiation gives
\begin{equation}
 \partial_x^2k_p=p^2k_p+\frac{p(2p+1)}2k_{p+1}.
 \label{eq:recursion_relation}
\end{equation}
The ratio $\mathcal N_{p+1}/\mathcal N_p=-p/(4\pi\gamma^2)$ therefore yields the normalized recurrence
\begin{equation}
 P(p+1;x)=\frac{p^2-\partial_x^2}{2\pi\gamma^2(2p+1)}P(p;x).
 \label{eq:dS_P_recursion}
\end{equation}
This compares response coefficients at the same formal $\lambda$; physically distinct couplings supply an additional ratio $\lambda_{p+1}^2/\lambda_p^2$. Appendix~\ref{app:dS_basecases} gives the $p=\tfrac12$ and $p=1$ bases. Apply all shift derivatives before setting $x=0$.

\subsection{Correlation terms}
\label{subsec:dS_X}
\label{subsec:dS_C}

For distinct static worldlines, $\kappa>1$. The Gaussian reduction gives
\begin{align}
 X(p;\kappa)&=\beta_1\int_0^\infty e^{-\beta_2s^2}\cos(\beta_3s)
 K_p^+(s;\kappa)\,ds,
 \label{eq:dS_X_integral}\\
 C(p;\kappa)&=\gamma_1\int_{\mathbb R}e^{-\gamma_2s^2-i\gamma_3s}
 K_p^-(s;\kappa)\,ds.
 \label{eq:dS_C_integral}
\end{align}
The distinct, explicitly normalized coefficient sets are given in Appendix~\ref{app:coefficients}. In particular, $\beta_1$ and $\gamma_1$ depend on $p$. The upper boundary in $X$ follows from $X=\rho_{14}$ and the ordering in Eq.~\eqref{eq:X_general}; complex switching coefficients are not conjugated along with the kernel.

A globally convergent Gegenbauer--Fourier expansion is unavailable here. Instead, use $\kappa$ as an auxiliary kernel parameter. Let $m=1$ for integer $p$, $m=\tfrac12$ for half-integer $p$, and $r=p-m\in\mathbb N_0$. Since $\partial_\kappa^rK_m^\pm=(-1)^r(m)_rK_p^\pm$, where $(m)_r=\Gamma(m+r)/\Gamma(m)$, the Gaussian prefactor ratio cancels the Pochhammer factor and gives
\begin{align}
 X(p;\kappa)&=\left(-\frac1{2\pi\gamma_A\gamma_B}\right)^r
 \partial_\kappa^rX(m;\kappa),
 \label{eq:dS_X_bootstrap}\\
 C(p;\kappa)&=\left(-\frac1{2\pi\gamma_A\gamma_B}\right)^r
 \partial_\kappa^rC(m;\kappa).
 \label{eq:dS_C_bootstrap}
\end{align}
These derivatives hold redshifts, switching parameters, and Gaussian coefficients fixed. The bases and convergence prescription appear in Appendix~\ref{app:dS_basecases}. At coincident worldlines, $C$ is defined by its original full-line distributional integral. The pointlike time-ordered $X$ has an ultraviolet divergence already at $\mathcal D=3$; the results for $X$ require positive spatial separation.

\subsection{Numerical checks and dimensional dependence}
\label{subsec:dS_numerical_check}

\begin{figure}[!tbp]
 \centering
 \includegraphics[width=0.95\columnwidth]{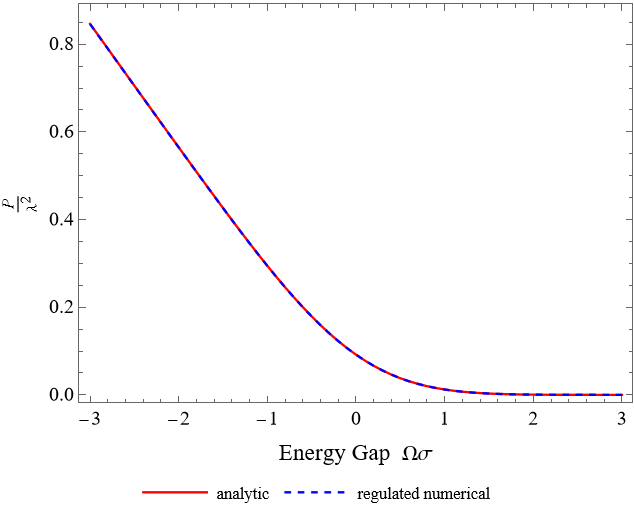}
 \caption{Gaussian-switched response $P/\lambda^2$ in $\mathrm{dS}_4$ at $R=0$, with $\alpha=\gamma=\sigma=1$. The solid curve uses Eq.~\eqref{eq:dS_P_erfc_series} with its asymptotic tail summed; the dashed curve is the independent contour quadrature of Eq.~\eqref{eq:dS_P_integral}.}
 \label{fig:dS4_probability_check}
\end{figure}

Fig. \ref{fig:dS4_probability_check} compares the convergent erfc series~\eqref{eq:dS_P_erfc_series} with an independent time-domain calculation. The latter displaces the proper-time contour to $u=v-i\delta$, $0<\delta<2\pi\gamma$, below the real axis. No pole is crossed, so this evaluates the original distributional response without a finite Wightman regulator. The largest observed absolute discrepancy on the sampled grid is $3.4\times10^{-16}$ for $P/\lambda^2$.

\begin{figure}[!tbp]
 \centering
 \includegraphics[width=0.95\columnwidth]{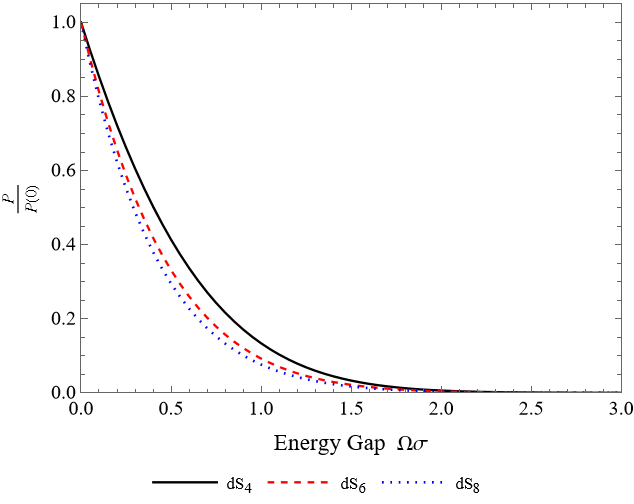}
 \caption{Normalized static-patch response $P_p(\Omega)/P_p(0)$ at $R=0$, $\alpha=\sigma=1$, in $\mathrm{dS}_4$, $\mathrm{dS}_6$, and $\mathrm{dS}_8$ ($p=1,2,3$). Curves use Eqs.~\eqref{eq:dS_spectral_recursion}--\eqref{eq:dS_spectral_P}, with separate time-contour checks.}
 \label{fig:dS_dimensional_probability}
\end{figure}

For an independent dimension check, define the stationary spectrum $\mathcal S_p(\omega)=\int_{\mathbb R}e^{-i\omega u}W_p(u)\,du$. Residues of the same-worldline hyperbolic kernels give the two base spectra, and the normalized kernel identity raises their dimension:
\begin{align}
 \mathcal S_{1/2}(\omega)&=\frac1{2(e^{2\pi\gamma\omega}+1)},\nonumber\\
 \mathcal S_1(\omega)&=\frac{\omega}{2\pi(e^{2\pi\gamma\omega}-1)},\nonumber\\
 \mathcal S_{p+1}(\omega)&=\frac{\omega^2+p^2/\gamma^2}{2\pi(2p+1)}\mathcal S_p(\omega),
 \label{eq:dS_spectral_recursion}\\
 P_p&=\lambda^2\sigma^2\int_{\mathbb R}e^{-\sigma^2(\Omega-\omega)^2}
 \mathcal S_p(\omega)\,d\omega.
 \label{eq:dS_spectral_P}
\end{align}
The removable value is $\mathcal S_1(0)=1/(4\pi^2\gamma)$. This positive spectrum has local temperature $(2\pi\gamma)^{-1}$. Fig. \ref{fig:dS_dimensional_probability} uses its Gaussian convolution to display the gap dependence for three choices of spacetime dimension; normalization by $P_p(0)$ removes the coupling and overall dimensional scale.

\subsection{Thermal restriction on the entangling gap}
\label{subsec:thermal_gap}

For thermal field states, Ref.~\cite{SimidzijaMartinMartinez2018} proves that harvesting decreases with temperature under broad switching and smearing assumptions. We hold the state and geometry fixed instead, and determine the entangling gaps and the maximum of $|X|/P$ for a Gaussian-switched pair.

Consider two distinct static detectors at the same radius $0<R<\alpha$, with the common width and frequency in Eq.~\eqref{eq:Gaussian_switching}. Their redshifts are equal, $\eta_A=\eta_B=\eta$, and $P_A=P_B=P$. The separation and lag $t_0$ are otherwise arbitrary. With $u=(\tau_A+\tau_B)/2$ and $s=\tau_A-\tau_B$, the switching product is
$\exp[-u^2/\sigma^2-(s+\eta t_0)^2/(4\sigma^2)]$.
The ordered kernel in Eq.~\eqref{eq:X_general} depends only on $s$, while its frequency phase is $e^{-2i\Omega u}$. Integrating $u$ therefore gives
\begin{equation}
 X(\Omega)=e^{-\sigma^2\Omega^2}X(0).
 \label{eq:Gaussian_X_factorization}
\end{equation}
This step uses stationarity and equal redshifts, rather than a particular field spectrum.

Write $\Xi=|X(0)|$ and $F(\Omega)=e^{\sigma^2\Omega^2}P(\Omega)$. Equation~\eqref{eq:dS_spectral_P} becomes
\begin{align}
 F(\Omega)&=\lambda^2\sigma^2\!\int_{\mathbb R}
 e^{-\sigma^2\omega^2+2\sigma^2\Omega\omega}
 \mathcal S_p(\omega)\,d\omega,\nonumber\\
 \mathcal C_2(\Omega)&=2e^{-\sigma^2\Omega^2}[\Xi-F(\Omega)]_+,
 \label{eq:Gaussian_noise_criterion}
\end{align}
where $[z]_+=\max(0,z)$ and $\mathcal C_2$ denotes the leading concurrence. The positive spectral weight makes $F$ strictly convex. In the sign convention of Eq.~\eqref{eq:dS_spectral_recursion}, the KMS relation is
$\mathcal S_p(-\omega)=e^{\beta\omega}\mathcal S_p(\omega)$, with $\beta=2\pi\gamma$.
Changing $\omega$ to $-\omega$ in the integral gives
\begin{equation}
 \begin{gathered}
 F(\Omega)=F(c-\Omega),\qquad c=\frac{\beta}{2\sigma^2},\\
 \Omega_*:=\underset{\Omega}{\operatorname{arg\,min}}F(\Omega)
 =\frac{\beta}{4\sigma^2}=\frac{\pi\gamma}{2\sigma^2}.
 \end{gathered}
 \label{eq:KMS_gap_reflection}
\end{equation}
Since $\Omega_*>0$, the existence test applies directly to physical excitation gaps:
\begin{equation}
 \begin{gathered}
 \exists\,\Omega>0:\ \mathcal C_2(\Omega)>0\\
 \Longleftrightarrow\quad
 |X(0)|>e^{\sigma^2\Omega_*^2}P(\Omega_*).
 \end{gathered}
 \label{eq:single_gap_existence_test}
\end{equation}
Only $X(0)$ and $P(\Omega_*)$ are needed for this test. For $\Xi>0$, $\Omega_*$ also uniquely maximizes $|X|/P$, independently of dimension, separation, and lag at fixed $\gamma$ and $\sigma$. If $\Xi=0$, the ratio vanishes at every gap.

For the matched stationary pairs and positive spectra assumed above, KMS detailed balance gives the reflection without a dS-specific kernel. Changing the geometry can change $X(0)$ and $P$, but at fixed $\beta$ and $\sigma$ it cannot move $\Omega_*$.

The spectrum has nonzero weight on both frequency half-lines, so $F$ tends to infinity at both ends of the signed-frequency axis. When nonempty, the entangling-gap set is consequently one bounded interval symmetric about $\Omega_*$; physical excitation gaps are its intersection with $\Omega>0$. The concurrence itself has a unique maximum at a smaller positive gap, $0<\Omega_{\rm peak}<\Omega_*$. Indeed, $F''>0$ makes $\log\mathcal C_2$ strictly concave wherever $\mathcal C_2>0$, and differentiation of Eq.~\eqref{eq:Gaussian_noise_criterion} places its maximum between zero and $\Omega_*$. The optimal ratio and optimal entanglement amount therefore answer different questions.

\begin{figure}[!htbp]
 \centering
 \includegraphics[width=0.95\columnwidth]{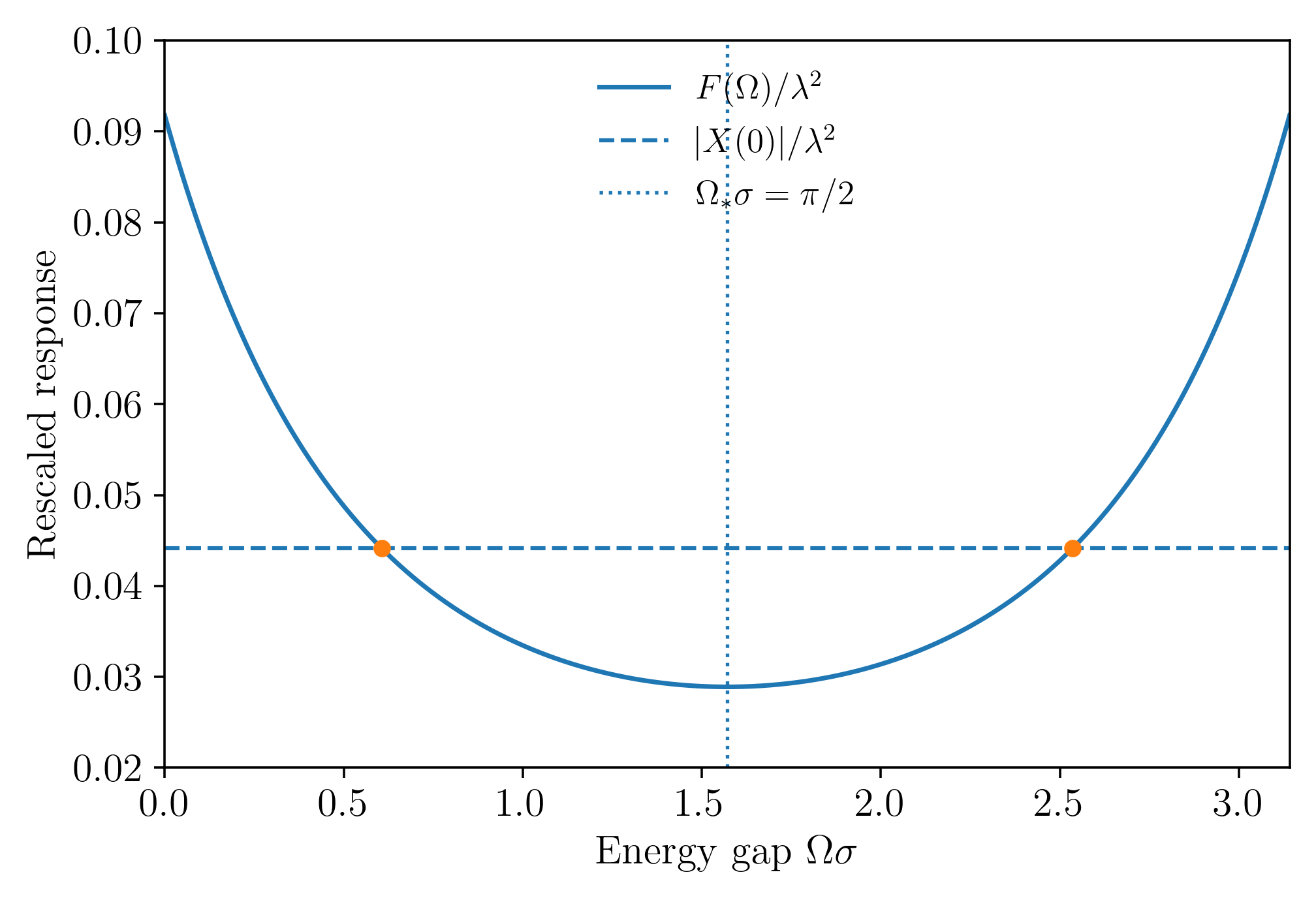}
 \caption{The single-gap criterion in $\mathrm{dS}_4$ for two static detectors with $R_A=R_B=\sigma$, $\alpha=\sqrt2\sigma$, angular separation $\Theta=\pi$, and $t_0=0$, so that $\gamma=\sigma$ and $\kappa=3$. The rescaled local response $F/\lambda^2=e^{\sigma^2\Omega^2}P/\lambda^2$ is compared with $\Xi/\lambda^2=|X(0)|/\lambda^2$. The vertical line marks $\Omega_*\sigma=\pi/2$; dots mark the two crossings. Leading concurrence is positive between them. The response is evaluated from Eq.~\eqref{eq:dS_spectral_P}, and $X(0)$ from the upper-boundary integral~\eqref{eq:dS_X_integral}; the Gaussian causal-overlap qualification remains in force.}
 \label{fig:dS_gap_criterion}
\end{figure}

The two crossings in Fig.~\ref{fig:dS_gap_criterion} bound the entangling interval for a pair at fixed positions and Gaussian width.

There is also a useful constraint on the origin of the correlations. Split the ordered kernel into its field anticommutator and commutator parts, giving $X=X_H+X_\Delta$. Under the same switching assumptions, $X_H$ is real, $X_\Delta$ is imaginary, and each obeys Eq.~\eqref{eq:Gaussian_X_factorization}. At zero gap, $X_H(0)=-\operatorname{Re}C(0)$, while positivity of the smeared two-point function gives $|C(0)|\le P(0)$. Convexity and $F(0)=F(c)$ then imply
\begin{equation}
 |X_H(\Omega)|>P(\Omega)
 \quad\Longrightarrow\quad 0<\Omega<\frac{\beta}{2\sigma^2}.
 \label{eq:KMS_symmetric_band}
\end{equation}
Outside this band, any positive total leading concurrence requires $|X_\Delta|^2>P^2-|X_H|^2\ge0$, since $|X|^2=|X_H|^2+|X_\Delta|^2$. This is a diagnostic of the perturbative criterion; the Gaussian interactions still have causal overlap.

The same Gaussian reduction applies to equal-redshift pairs in the global AdS and Minkowski vacua. Their spectra have support only at negative frequency, which gives $F'<0$, $F(+\infty)=0$, and $F(-\infty)=\infty$. For $\Xi>0$, the leading entangling-gap set is then a half-line instead of a bounded interval. Although the ratio improves indefinitely, the concurrence tends to zero as $\Omega\to+\infty$ by Eq.~\eqref{eq:Gaussian_noise_criterion}. These statements concern the leading perturbative coefficients and do not give a uniform large-gap estimate for the full detector state.

\section{Radial infall in the BTZ black hole}
\label{sec:BTZ}

The final application is to one detector in radial free fall, rather than a harvesting pair. The BTZ Hartle--Hawking function is an image sum over covering AdS, so the two responses can be compared at the same local curvature. Hodgkinson and Louko derived the three-dimensional detector prescription and image-integral responses for stationary and freely falling BTZ detectors \cite{HodgkinsonLouko2012}. Preciado-Rivas \emph{et al.} followed infall through the horizon numerically, finding finite-time glitches and rapid growth near the singularity \cite{PhysRevD.110.025002}.

We integrate those image kernels into a Legendre series. A uniform estimate of the image sum then gives the logarithmic endpoint law and its coefficient. We also smooth the onset to separate its finite-time glitches from the final growth.

\subsection{Wightman function}
\label{subsec:BTZ_Wightman}
The nonrotating BTZ metric is
\begin{align}
 ds^2&=-\frac{r^2-r_h^2}{\ell^2}dt^2
       +\frac{\ell^2}{r^2-r_h^2}dr^2+r^2d\phi^2,
\end{align}
with $r_h=\ell\sqrt M$, $M>0$, $\ell>0$, and $\phi\sim\phi+2\pi$.
For the massless
conformally coupled scalar, $(\Box-\mathcal R/8)\phi=0$ with
$\mathcal R=-6/\ell^2$, the Hartle--Hawking two-point function is
\cite{HodgkinsonLouko2012,PhysRevD.110.025002}
\begin{align}
 W_{\rm BTZ}(x,x')
 &=\frac{1}{4\pi\sqrt2\ell}\sum_{n\in\mathbb Z}
 \left[\sigma_{n,\epsilon}^{-1/2}
       -\zeta(\sigma_{n,\epsilon}+2)^{-1/2}\right],
\label{eq:BTZ_Wightman_images}
\end{align}
where the $\epsilon\downarrow0$ distributional limit is understood.
For exterior points $r,r'>r_h$,
\begin{align}
\begin{split}
 \sigma_{n,\epsilon}
 &=\frac{rr'}{r_h^2}
   \cosh\!\left[\frac{r_h}{\ell}(\Delta\phi-2\pi n)\right]-1\\
 &\quad-\frac{\sqrt{(r^2-r_h^2)(r'^2-r_h^2)}}{r_h^2}
   \cosh\!\left(\frac{r_h\Delta t}{\ell^2}-i\epsilon\right).
\end{split}
\label{eq:BTZ_Sigma_neps}
\end{align}
Here $\Delta t=t-t'$, $\Delta\phi=\phi-\phi'$, and the images shift
$\phi'\mapsto\phi'+2\pi n$. The parameter
$\zeta=-1,0,1$ denotes Neumann, transparent, and Dirichlet boundary
conditions, respectively.

\subsection{Sharply switched transition rates}
\label{sec:sharp_switching}

Let $\tau_0$ and $\tau>\tau_0$ be the proper times of switch-on and sharp readout. The rate below uses the three-dimensional Hadamard prescription \cite{HodgkinsonLouko2012}. The sharp-switching limit is dimension dependent.

With the regulator still present, let $P_{D,\epsilon}(\tau)$ be
Eq.~\eqref{eq:PD_general} with $W$ replaced by $W_\epsilon$ and both
proper-time integrals restricted to $(-\infty,\tau]$. At finite
$\epsilon>0$, differentiation gives
\begin{align}
\begin{split}
\frac{dP_{D,\epsilon}}{d\tau}
 &=2\lambda^2\chi(\tau)\int_0^\infty ds\,\chi(\tau-s)\\
 &\quad\times\Re\!\left[e^{-i\Omega s}W_\epsilon(\tau,\tau-s)\right].
\end{split}
\label{eq:transition_rate_general}
\end{align}
A smooth $\chi$ still acquires a sharp readout cutoff if $\chi(\tau)\ne0$. Equation~\eqref{eq:transition_rate_general} holds with the regulator present; it does not license pointwise regulator removal in arbitrary dimension.

For a Hadamard state in $2+1$ dimensions, first extracting the coincidence
contribution and then taking the controlled sharp-switching limit gives
\cite{HodgkinsonLouko2012}
\begin{align}\label{eq:sharp_switching_rate}
    \dot P_D(\tau)
=\lambda^2\Bigg[\frac14+2\int_0^{\Delta\tau}ds\times\Re\!\left(e^{-i\Omega s}W_0(\tau,\tau-s)\right)\Bigg],
\end{align}
where $\Delta\tau=\tau-\tau_0$, the dot means proper-time differentiation,
and $W_0$ is the noncoincident boundary value for $s>0$ \emph{after} the
contact term has been separated. Null singularities away from coincidence are assumed integrable, as in BTZ. The universal $1/4$ is the contact contribution in three dimensions; retaining the coincidence distribution in $W_0$ would double-count it. For example, an inertial detector in the
massless Minkowski vacuum has
$W_0(s)=-i/(4\pi s)$ and
$\dot P_D/\lambda^2=1/4-\operatorname{Si}(\Omega\Delta\tau)/(2\pi)$,
where $\operatorname{Si}(x)=\int_0^x(\sin u)/u\,du$.

\subsection{Transition rate of a radially infalling detector}
\label{subsec:BTZ_infall_rate}
Use dimensionless proper time $\tau=\tau_{\rm phys}/\ell$ and signed
frequency $E=\ell E_{\rm phys}$. A radial geodesic released from rest at
$r=\varrho r_h$, with $\varrho>1$, has
\begin{align}
 r(\tau)&=\varrho r_h\cos\tau, \,\, \phi(\tau)=\phi_0,\\
 t(\tau)&=\frac{\ell}{\sqrt M}
 \operatorname{arctanh}\!\left(\frac{\tan\tau}{\sqrt{\varrho^2-1}}\right).
\end{align}
The displayed $t$ coordinate is real only in the exterior. The geodesic
and its invariant two-point pullback continue regularly across the
horizon at $\tau_h=\arccos(1/\varrho)$; the latter, rather than a numerical
continuation of $t$, is used below. We take the detector to switch on at
release, $\tau_0=0$, so $\Delta\tau=\tau\in(0,\pi/2)$, ending before
the singularity. All integration times $s$ below are also dimensionless.

Substitution into Eq.~\eqref{eq:sharp_switching_rate} yields
\begin{align}
 \dot{\mathcal F}_\tau(E)&=\frac14+\sum_{n\in\mathbb Z}A_n(\tau;E),
\label{eq:BTZ_infall_rate_imagesum}\\
 \frac{dP}{d\tau_{\rm phys}}&=\lambda_{\rm phys}^2\dot{\mathcal F}_\tau,\quad
 \frac{dP}{d\tau}=\bar\lambda^2\dot{\mathcal F}_\tau.
\end{align}
Thus $\dot{\mathcal F}_\tau$ is dimensionless and
$[\lambda_{\rm phys}^2]=L^{-1}$; equivalently the dimensionless-time
coupling is $\bar\lambda=\sqrt\ell\lambda_{\rm phys}$.
The image contribution is
\begin{align}
\begin{split}
 A_n&=\frac{1}{2\pi\sqrt2}\Re\int_0^{\Delta\tau}ds\,e^{-iEs}\\
 &\quad\times\left[(D_n^-)^{-1/2}-\zeta(D_n^+)^{-1/2}\right],
\end{split}
\label{eq:BTZ_infall_An_start}\\
 D_n^\pm&=\pm1+K_n\cos\tau\cos(\tau-s)\nonumber\\
 &\quad+\sin\tau\sin(\tau-s),\nonumber\\
 K_n&=1+2\varrho^2\sinh^2(n\pi\sqrt M).
\label{eq:BTZ_Kn_def}
\end{align}
The square roots inherit the continuation $s\mapsto s-i0$.

For $n\ge1$, define
\begin{align}
 L_n&=1+(K_n-1)\cos^2\tau,\nonumber\\
 Q_n&=(K_n-1)\sin\tau\cos\tau,\nonumber\\
 R_n&=\sqrt{L_n^2+Q_n^2},\quad
 \varphi_n=\arctan(Q_n/L_n),\nonumber\\
 \alpha_n&=R_n^{-1},\qquad\theta_n=\arccos\alpha_n.
\label{eq:BTZ_phase_parameters}
\end{align}
Then $D_n^\pm=R_n[\cos(s-\varphi_n)\pm\alpha_n]$ and
$0<\alpha_n<1$, since
$R_n^2=1+(K_n^2-1)\cos^2\tau>1$.
The boundary denominator $D_n^+$ stays positive. The direct denominator
starts positive and has at most one root in the integration interval,
$s_n^*=\varphi_n+\theta_n$. When this root is present, its square root
is positive before it and positive imaginary after it. The real direct
integrand in Eq.~\eqref{eq:BTZ_infall_An_start} is consequently
$\cos(Es)/\sqrt{D_n^-}$ for $s<s_n^*$ and
$-\sin(Es)/\sqrt{-D_n^-}$ for $s>s_n^*$.
The inverse-square-root singularity is integrable.

With $u=s-\varphi_n$, the normalized shifted expression is
\begin{align}
\begin{split}
 A_n={}&\frac{\sqrt{\alpha_n}}{2\pi\sqrt2}\Re\Bigg\{
 e^{-iE\varphi_n}\int_{-\varphi_n}^{\Delta\tau-\varphi_n}du\,e^{-iEu}\\
 &\times\Bigl[(\cos(u-i0)-\alpha_n)^{-1/2}\\
 &\hspace{3em}-\zeta(\cos(u-i0)+\alpha_n)^{-1/2}\Bigr]\Bigg\}.
\end{split}
\label{eq:BTZ_shifted_kernel}
\end{align}
The Legendre generating function, first at $\eta>0$, gives
\begin{multline}
 [\cos(u-i\eta)\mp\alpha]^{-1/2}\\
 =\sqrt2\sum_{m=0}^\infty P_m(\pm\alpha)
 e^{-i(m+1/2)(u-i\eta)}.
\end{multline}
Integrating before taking $\eta\downarrow0$ yields
\begin{align}
 A_n={}&\frac{\sqrt{\alpha_n}}{2\pi}\sum_{m=0}^\infty
 [P_m(\alpha_n)-\zeta P_m(-\alpha_n)]\,S_m,
\label{eq:BTZ_infall_An_double_series}\\
 S_m={}&\Delta\tau\,\operatorname{sinc}\!\left(
 \frac{(E+\omega_m)\Delta\tau}{2}\right)\nonumber\\
 &\times\cos\!\left(\frac{(E+\omega_m)\Delta\tau}{2}
             -\omega_m\varphi_n\right),\nonumber
\end{align}
where $\omega_m=m+1/2$ and $\operatorname{sinc}x=\sin x/x$, with
$\operatorname{sinc}0=1$. This evaluates the elementary integral
\begin{align}
 S_m=\int_0^{\Delta\tau}ds\,
 \cos[(E+\omega_m)s-\omega_m\varphi_n].
\end{align}
At $E=-\omega_m$ it equals $\Delta\tau\cos(\omega_m\varphi_n)$;
there is no resonant pole.

For fixed interior trajectory parameters,
$P_m(\pm\alpha_n)=O(m^{-1/2})$ and $S_m=O(m^{-1})$.
The integrated mode series is therefore absolutely convergent, including
at a null endpoint, although the unevaluated Fourier series requires
its Abel boundary value. Also
$\sqrt{\alpha_n}=O(e^{-n\pi\sqrt M})$, giving absolute convergence of
the image sum on compact portions of the trajectory. The mode
convergence is algebraic, not uniformly rapid near glitches;
neither $M\to0$ nor $\tau\to\pi/2$ is a uniform image-truncation limit.

Since $K_{-n}=K_n$, the full result is
$\dot{\mathcal F}_\tau=\dot{\mathcal F}^{(0)}_\tau
+2\sum_{n\ge1}A_n$, with the separately evaluated zero image
\begin{align}
 \dot{\mathcal F}^{(0)}_\tau(E)
 =\frac14-\frac{1}{4\pi}\int_0^{\Delta\tau}ds
 \times\left[\frac{\sin(Es)}{\sin(s/2)}
       +\zeta\frac{\cos(Es)}{\cos(s/2)}\right].
\label{eq:BTZ_infall_n0_term}
\end{align}
This regular finite integral is valid for $0<\Delta\tau<\pi$;
the first ratio has limit $2E$ at zero. It is the pure AdS$_3$
contribution. Although the Legendre generating function exists at
$\alpha_0=1$ as an Abel expansion, the preceding integrated endpoint
argument does not apply there. Eqs. \eqref{eq:BTZ_infall_An_double_series}
and~\eqref{eq:BTZ_infall_n0_term} give an explicit mode representation
plus one finite integral, elementary for integer $E$.

\subsection{Image accumulation near the singularity}
\label{subsec:BTZ_endpoint}
The covering-AdS contribution in Eq.~\eqref{eq:BTZ_infall_n0_term}
stays finite as the infaller approaches $r=0$. The full BTZ rate has a
different limit. Set $T=\pi/2$, $\delta=T-\tau$, and $a=\pi\sqrt M$.
For fixed $M>0$, $\varrho>1$, $E\in\mathbb R$, and
$\zeta\in\{-1,0,1\}$, we find
\begin{align}
 \dot{\mathcal F}_\tau(E)
 &=\frac{B_\zeta(E)}{a}\log\frac1\delta+O(1),
 \label{eq:BTZ_endpoint_log}\\
 B_\zeta(E)&=\frac14+\dot{\mathcal F}^{(0)}_T(E)>0.
 \label{eq:BTZ_endpoint_B}
\end{align}
Thus the zero-image integral also determines the coefficient of the
divergence. The leading coefficient is independent of the release radius
$\varrho$; the bounded remainder need not be. For example,
$B_\zeta(0)=\tfrac12-\tfrac{\zeta}{2\pi}\log(1+\sqrt2)$.
Appendix~\ref{app:BTZ_endpoint} gives a uniform estimate of the image sum;
taking the singular limit image by image would miss a finite
contribution from a shrinking interval near $s=0$.

The positive-index image band in Eq.~\eqref{eq:BTZ_image_count} contains $(2a)^{-1}\log(1/\delta)+O(1)$ terms. Pairing positive and negative images gives the coefficient in Eq.~\eqref{eq:BTZ_endpoint_log}. The divergent rate therefore comes from image accumulation, not from the covering-AdS term.

The logarithm persists when the switch-on is smoothed. To make this
statement precise, let $\chi\in C^1([0,T];\mathbb R)$ be fixed while the
readout time varies, and restrict the interaction to $0\le u\le\tau$.
Write $\chi_*=\chi(T)$ and
$\mathscr R_\chi=\bar\lambda^{-2}dP_\chi/d\tau$ at leading order. The readout endpoint remains sharp. The coefficient generally depends on the full onset profile.
Then
\begin{align}
 \mathscr R_\chi(\tau;E)
 &=\frac{B_{\chi,\zeta}(E)}{a}\log\frac1\delta+O(1),
 \label{eq:BTZ_smooth_log}\\
 B_{\chi,\zeta}(E)
 &=\frac{\chi_*^2}{2}
 -\frac{\chi_*}{4\pi}\int_0^T ds\,\chi(T-s)
 \nonumber\\[-1ex]
 &\qquad\times\left[
 \frac{\sin(Es)}{\sin(s/2)}
 +\zeta\frac{\cos(Es)}{\cos(s/2)}\right].
 \label{eq:BTZ_smooth_B}
\end{align}
For a nonnegative, nondecreasing onset with $\chi_*>0$, this coefficient
is strictly positive for every signed $E$ and every boundary condition.
If the extension of $\chi$ by zero before release is $C^\infty$, the
finite-time onset glitches disappear, while
Eq.~\eqref{eq:BTZ_smooth_log} still holds. At $E=0$ with transparent
boundary conditions, $B_{\chi,0}(0)=\chi_*^2/2$, independently of the
onset history.

The divergent rate is integrable. The final interval contributes
$\int_{T-\delta}^{T}\mathscr R_\chi(u;E)\,du
=(B_{\chi,\zeta}/a)\delta\log(1/\delta)+O(\delta)$,
so the accumulated leading-order response has a finite endpoint limit.
Here the parameters and onset profile are fixed. There is no uniform
$M\to0$ limit or continuation through the singularity.

\subsection{Numerical comparisons: finite-time features and endpoint growth}
\label{subsec:BTZ_numerical}

\begin{figure*}[t]
 \centering
 \includegraphics[width=0.95\textwidth]{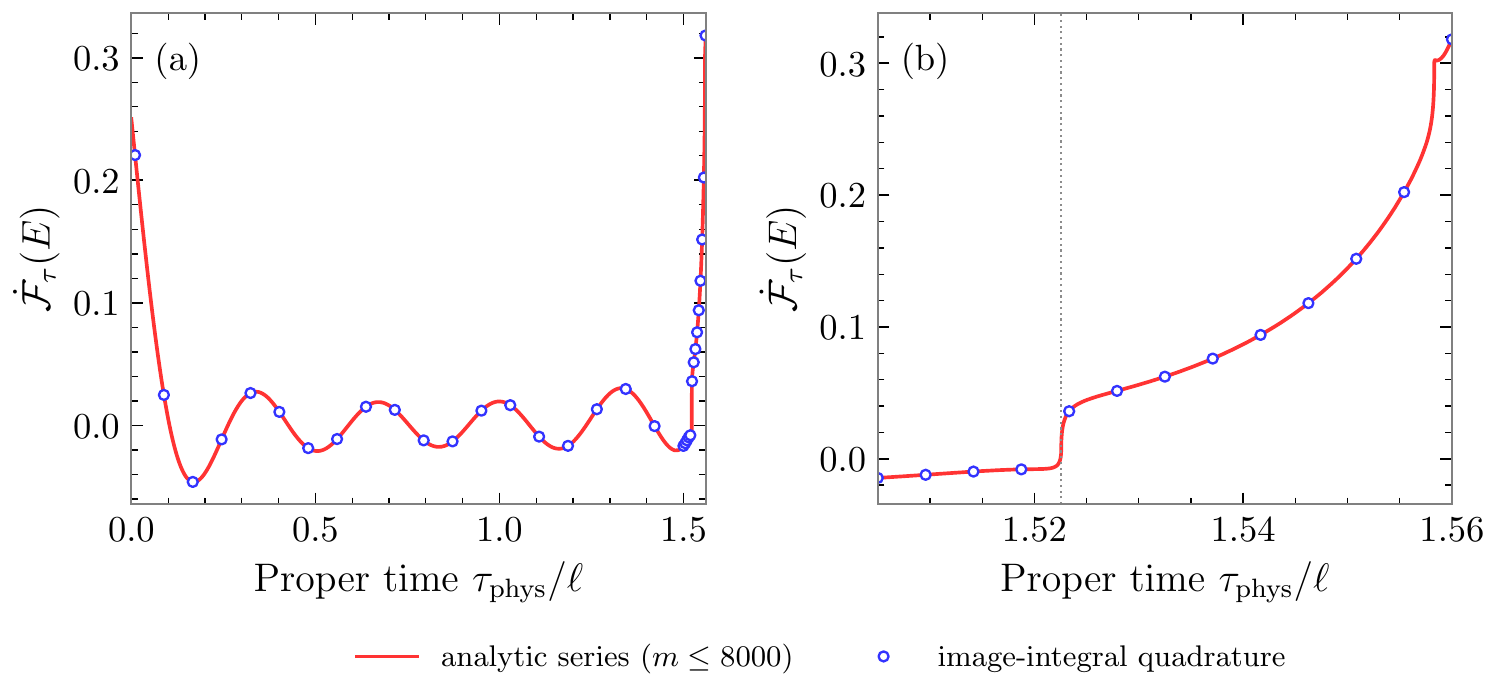}
 \caption{Sharp BTZ response for $M=10^{-4}$, $\varrho=100$, $E=20$,
 $\zeta=0$, and release switch-on. The series uses $m_{\max}=8000$,
 $n_{\max}=600$, and the elementary zero image; circles are independent
 image-integral quadratures. The largest observed discrepancy at the
 plotted markers is $1.16\times10^{-4}$. Panel (b) enlarges the first
 glitch, $\tau_1\simeq1.522575$ (dotted line). The displayed range ends
 at $1.56$, just before the horizon $\tau_h\simeq1.560796$.
 A negative instantaneous rate does not imply a negative probability.}
 \label{fig:BTZ_infall_check}
\end{figure*}

Figure~\ref{fig:BTZ_infall_check} compares the integrated modes with independent image-integral quadrature at the parameters of Ref.~\cite{PhysRevD.110.025002}. Its range ends before horizon crossing and does not test the endpoint logarithm. We split the quadrature at $s_n^*$ and use $s=s_n^*\mp x^2$ to remove the square-root endpoints. Release-switching glitches occur at $\tau_n=\arccos(1/K_n)$, when $s_n^*=\Delta\tau$. The rate is finite but generically not differentiable there; these are not horizon crossings. Appendix~\ref{app:BTZ_numerical_details} records the truncation and interpolation checks.

\begin{figure}[!htbp]
 \centering
 \includegraphics[width=0.95\columnwidth]{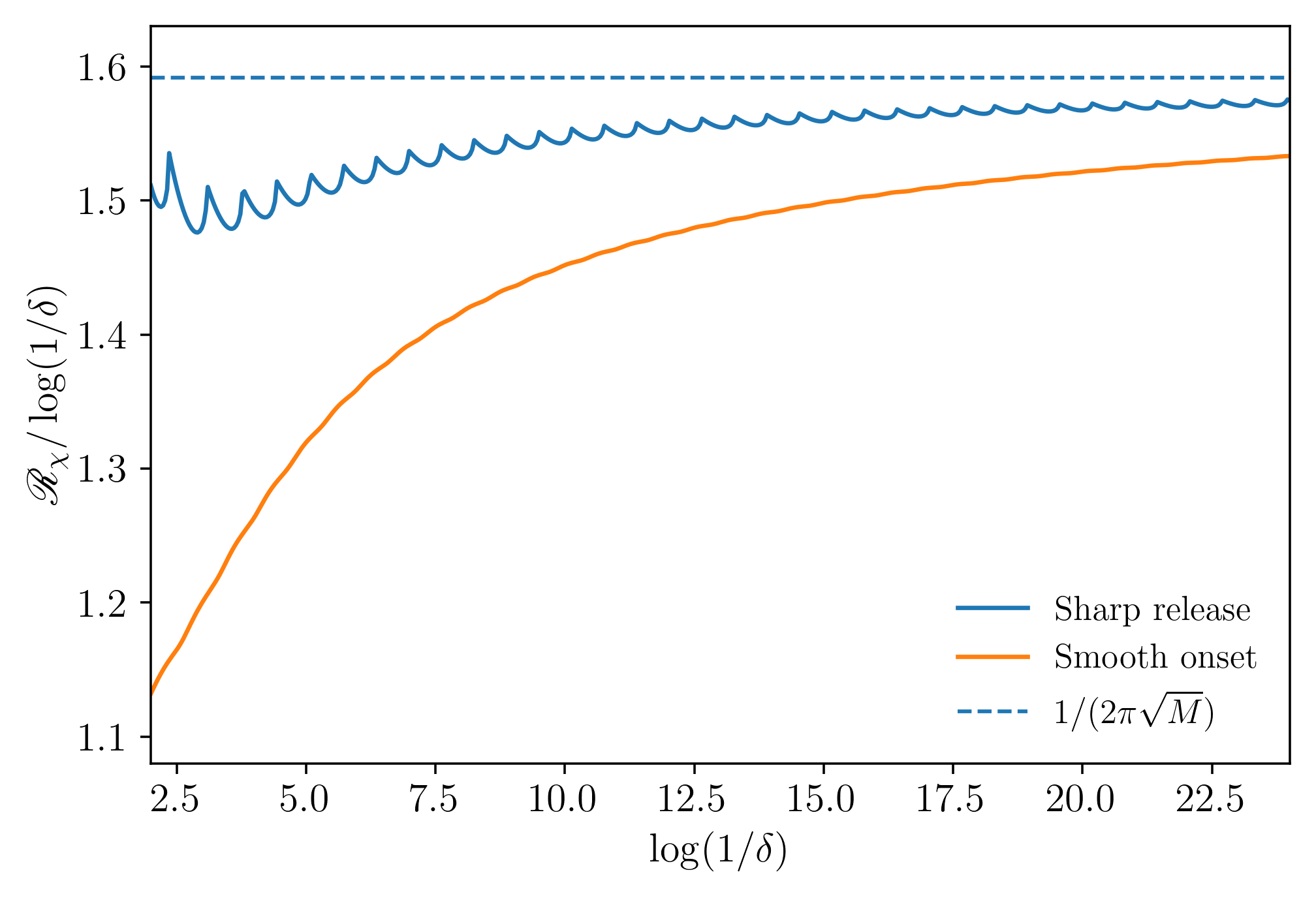}
 \caption{BTZ endpoint growth at $M=10^{-2}$, $\varrho=2$, $E=0$, and $\zeta=0$. We plot $\mathscr R_\chi/\log(1/\delta)$, with $\delta=\pi/2-\tau$, for sharp release and the smooth onset above. The horizontal line is the predicted limit $1/(2\pi\sqrt M)$, not a fitted value. Weighted image integrals are evaluated in the root-removing coordinate~\eqref{eq:BTZ_endpoint_factor}. The readout endpoint remains sharp.}
 \label{fig:BTZ_endpoint_comparison}
\end{figure}

Figure~\ref{fig:BTZ_endpoint_comparison} compares sharp release with $\chi_{\rm sm}(u)=h(u/u_s)$ at $E=0$, $\zeta=0$. Here $u_s=1.3$ and $h(x)=f(x)/[f(x)+f(1-x)]$. We take $f(x)=e^{-1/x}$ for $x>0$ and zero otherwise. The onset is nonnegative, nondecreasing, and smooth across release and its plateau. Both profiles have $\chi_*=1$, so Eq.~\eqref{eq:BTZ_smooth_B} gives $B_{\chi,0}(0)=1/2$ for each. Their finite-time responses differ, but their logarithmic coefficients agree. This equality need not hold at other gaps.

\section{Concluding remarks}
\label{sec:conclusion}

We have developed analytic expressions for UDW
detector responses and correlations across several constant-curvature
spacetimes, including Minkowski, AdS, dS, and BTZ geometries. For matched, equal-redshift Gaussian pairs with equal local responses, the field spectrum constrains which gaps can produce entanglement. In the dS static patch, only $X(0)$ and $P(\Omega_*)$ are needed to test its existence, with $\Omega_*=\beta/(4\sigma^2)$. A nonempty entangling interval is bounded and symmetric about $\Omega_*$ on the signed-frequency axis. This gap maximizes $|X|/P$; the concurrence peaks at a smaller positive gap. In the Minkowski and global AdS vacua, $X(0)\ne0$ instead gives a half-line of leading entangling gaps. The concurrence tends to zero at large gap despite the absence of an upper threshold.

In BTZ, the bounded covering-AdS response is replaced by a logarithmic endpoint rate. A uniform estimate shows how the additional quotient images accumulate and determines the coefficient. It is positive for nonnegative, nondecreasing onsets with nonzero terminal amplitude. An onset whose zero extension is smooth removes the finite-time glitches but leaves this final growth. The readout endpoint remains sharp. Since the logarithm is integrable, the accumulated leading-order response stays finite.

Table~\ref{tab:analytic_lookup} collects the response formulas and their domains. Their series have different tails and require separate truncation estimates; analytic evaluation alone implies no numerical speed advantage. A fixed BTZ image cutoff does not give a uniform approximation at the endpoint or as $M\to0$. Other switchings require their own Fourier and half-Fourier transforms. Coincident pointlike $X$ still needs ultraviolet input. Our entanglement statements concern leading perturbative coefficients, not the full state uniformly at large gap. Gaussian interactions retain causal overlap, so positive concurrence need not be a signaling-free harvesting result.

More broadly, our analytic results illustrate explicitly how localized quantum probes can serve as operational diagnostics of quantum fields in curved spacetime. Analytic detector models provide a controlled setting for connecting observables obtained from local quantum measurements to global properties of spacetime and quantum field states. We expect that the formulas derived and methods used here will  be useful beyond the specific examples considered, both as
benchmarks for numerical studies and as a guide for identifying which
features of a detector signal originate from the global structure of
spacetime, the quantum state, or the interaction protocol. Taken together, these results highlight the broader potential of localized
quantum measurements for probing how global properties of spacetime and
quantum fields are encoded in observable detector responses.

\begin{acknowledgments}
This work was supported in part by the Natural Sciences and Engineering Research Council of Canada. J.Y. acknowledges support from the Natural Sciences and Engineering Research Council of Canada through a Canada Graduate Research Scholarship--Doctoral [Funding Reference Number: \texttt{110\_2026\_2027\_Q1\_571}]. M.Z. was supported by the China Scholarship Council Postdoctoral Fellowship (No. 202208360067) and the National Natural Science Foundation of China (Grant No. 12365010). Research at Perimeter Institute is supported in part by the Government of Canada through the Department of Innovation, Science, and Economic Development and by the Province of Ontario through the Ministry of Colleges and Universities.
\end{acknowledgments}

\appendix
\clearpage
\onecolumngrid

\section{Minkowski correlation integrals}
\label{app:Mink_C}

For $p>0$, $L>0$, $a>0$, and $b\in\mathbb C$, the kernel $C_{\mathrm M}$ of Eq.~\eqref{eq:Mink_C_integral} is holomorphic in $b$. The Laplace transform of $k^{p-1/2}J_{p-1/2}(Lk)$ gives, at finite $\epsilon>0$,
\begin{align}
 [L^2-(s-i\epsilon)^2]^{-p}
 &=\int_0^\infty e^{-iks-\epsilon k}\widetilde D_p(k;L)\,\mathrm dk,
 \label{eq:Mink_C_fourier_rep}\\
 \widetilde D_p(k;L)
 &=\frac{\sqrt\pi}{\Gamma(p)}
 \left(\frac{k}{2L}\right)^{p-1/2}J_{p-1/2}(Lk).
 \label{eq:Mink_C_D_general}
\end{align}
The damping licenses interchange of the integrals. Gaussian integration followed by dominated convergence as $\epsilon\to0^+$ yields
\begin{align}
 C_{\mathrm M}(p;L;a,b)
 &=\sqrt{\frac\pi a}\int_0^\infty
 \widetilde D_p(k;L)e^{-(k+b)^2/(4a)}\,\mathrm dk.
 \label{eq:Mink_C_master}
\end{align}
This integral is absolutely convergent, including for the complex $b$ in Eq.~\eqref{eq:Mink_C_parameters}. Equivalently, the physical coherence is
\begin{align}
 C(t_0)=2\pi\lambda^2\mathcal A_p\sigma^2
 \int_0^\infty\widetilde D_p(k;L)
 e^{-\sigma^2(k+\Omega)^2+i(k+\Omega)t_0}\,\mathrm dk.
 \label{eq:Mink_C_physical_spectral}
\end{align}
The limit $\widetilde D_p(k;0)=k^{2p-1}/\Gamma(2p)$ proves $C=P$ at $L=t_0=0$.

For $p=1/2$, $\widetilde D_{1/2}=J_0(Lk)$, hence
\begin{align}
 C_{\mathrm M}(\tfrac12;L;a,b)
 &=\sqrt{\frac\pi a}\int_0^\infty J_0(Lk)e^{-(k+b)^2/(4a)}\,\mathrm dk,
 \label{eq:Mink_C_half}\\
 C_{\mathrm M}(\tfrac12;L;a,0)
 &=\pi e^{-aL^2/2}I_0(aL^2/2).
 \label{eq:Mink_C_half_centered}
\end{align}
For $p=1$, $\widetilde D_1=\sin(Lk)/L$. Define the entire function of $b$
\begin{align}
 F(b,L)&=e^{-aL^2-ibL}\erfc\!\left(\frac b{2\sqrt a}-iL\sqrt a\right).
 \label{eq:Mink_C_F}
\end{align}
Completing the Gaussian square separately for $e^{iLk}$ and $e^{-iLk}$ gives
\begin{align}
 C_{\mathrm M}(1;L;a,b)
 &=\frac\pi{2iL}\bigl[F(b,L)-F(b,-L)\bigr],
 \label{eq:Mink_C_one}\\
 C_{\mathrm M}(1;L;a,0)
 &=\frac\pi L e^{-aL^2}\erfi(L\sqrt a).
 \label{eq:Mink_C_one_centered}
\end{align}
Only when $b$ is real may the difference be written as $(\pi/L)\Im F(b,L)$; Eq.~\eqref{eq:Mink_C_one} is the required extension to nonzero lag. Differentiating the boundary distribution at fixed $a,b$ gives
\begin{align}
 C_{\mathrm M}(p+1;L;a,b)
 &=-\frac1p\frac{\partial C_{\mathrm M}(p;L;a,b)}{\partial L^2}.
 \label{eq:Mink_C_raising}
\end{align}
These two bases therefore determine all integer and half-integer $p>0$. For the normalized physical coefficient the corresponding rule is $C_{p+1}=-(1/\pi)\partial_{L^2}C_p$, with the same formal coupling and other parameters fixed.

For the independent numerical checks, shift the full-line $P$ and $C$ contours to $s=u-i\delta$, $\delta>0$, in their Gaussian-smeared Wightman integrals. The kernels are holomorphic below the real axis and Gaussian decay removes the vertical segments at infinity. For $X$, put
$f(s)=e^{-s^2/(4\sigma^2)}\cosh[st_0/(2\sigma^2)]$ and deform into the upper half-plane:
\begin{align}
 \int_0^\infty\frac{f(s)\,\mathrm ds}{[L^2-(s+i0)^2]^p}
 &=i\int_0^\delta\frac{f(iv)\,\mathrm dv}{(L^2+v^2)^p}
 +\int_0^\infty\frac{f(u+i\delta)\,\mathrm du}{[L^2-(u+i\delta)^2]^p}.
 \label{eq:Mink_X_contour}
\end{align}
The finite vertical segment is essential. These identities evaluate the boundary distributions exactly; numerical error comes from quadrature and roundoff, rather than from a retained regulator.

\section{Coefficient definitions}
\label{app:coefficients}
The Gaussian integrations for static AdS and dS pairs share a geometric prefactor but have distinct frequency and lag coefficients for $X$ and $C$. Set $a=\gamma_A$, $b=\gamma_B$, $S=a^2+b^2$, and $\delta=t_0/L_c$, where $L_c=\ell$ in AdS and $L_c=\alpha$ in dS. Here $\gamma_D=\sqrt{\ell^2+R_D^2}$ in AdS and $\gamma_D=\sqrt{\alpha^2-R_D^2}$ in dS. Define
\begin{equation}\label{eq:Gaussian_common}
A=\frac{a^2b^2}{2\sigma^2S},\qquad
Q_p=\frac{\lambda^2\Gamma(p)}{2^{p+2}\pi^{p+1}}
(ab)^{1-p}\frac{\sqrt{2\pi}\sigma}{\sqrt S}.
\end{equation}
Completing the square in the second integration variable gives
\begin{align}\label{eq:Gaussian_X_coefficients}
B_X&=\frac{\Omega ab(a-b)}S+2iA\delta,&
E_X&=\frac{\sigma^2\Omega^2(a+b)^2}{2S}+A\delta^2
+\frac{i\Omega\delta(b-a)(a+b)^2}{2S},\\
\label{eq:Gaussian_C_coefficients}
B_C&=\frac{\Omega ab(a+b)}S-2iA\delta,&
E_C&=\frac{\sigma^2\Omega^2(a-b)^2}{2S}+A\delta^2
-\frac{i\Omega\delta(a-b)^2(a+b)}{2S}.
\end{align}
For the difference variable $s=(t_A-t_B)/L_c$, the full-line $C$ test function is $Q_pe^{-E_C}e^{-As^2-iB_Cs}$. The two ordered halves of $X$ instead sum to $-2Q_pe^{-E_X}e^{-As^2}\cos(B_Xs)$ on $s>0$, multiplying the upper-boundary kernel. Hence the coefficient dictionaries used in the main text are
\begin{equation}\label{eq:coefficient_dictionary}
\begin{aligned}
&(a_1,a_2,a_3)=(-2Q_pe^{-E_X},A,B_X) &&\text{(AdS $X$)},\\
&(\beta_1,\beta_2,\beta_3)=(-2Q_pe^{-E_X},A,B_X) &&\text{(dS $X$)},\\
&(\gamma_1,\gamma_2,\gamma_3)=(Q_pe^{-E_C},A,B_C) &&\text{(dS $C$)}.
\end{aligned}
\end{equation}
Both $a_1,\beta_1$ and $\gamma_1$ depend on $p$ and contain $\lambda^2$. The separation derivative used in dS dimension raising acts on the kernel parameter $\kappa$ while all entries in this dictionary and $a,b$ are held fixed.

\section{Subtraction data and Abel sums for \texorpdfstring{$\mathrm{AdS}_{d+1}$}{AdS}}
\label{app:AdS_sums}

For $f(s)=e^{-a_2s^2}\cos(a_3s)$, the derivatives required in Eq.~\eqref{eq:AdS_X0_final} are
\begin{equation}\label{eq:AdS_f_derivs}
 f^{(2j)}(0)=(-1)^j(2j)!\sum_{k=0}^{j}
 \frac{a_2^{j-k}a_3^{2k}}{(j-k)!(2k)!}.
\end{equation}
The restoration sums are defined, for $-1\le\alpha<1$, by
\begin{equation}
 S_{2j+1}(\alpha)=\lim_{\epsilon\downarrow0}
 \sum_{n=0}^{\infty}\frac{C_n^p(\alpha)e^{-(n+p)\epsilon}}{(n+p)^{2j+1}}.
\end{equation}
Applying the Mellin representation of $(n+p)^{-2j-1}$ at positive $\epsilon$ and then the generating function gives an integral with $(\cosh(t+\epsilon)-\alpha)^{-p}$. Dominated convergence therefore yields
\begin{equation}\label{eq:AdS_Sodd_integral}
 S_{2j+1}(\alpha)=\frac{2^{-p}}{(2j)!}
 \int_0^\infty dt\,t^{2j}(\cosh t-\alpha)^{-p}.
\end{equation}
This integral is finite also at $\alpha=-1$. At $\alpha=1$, however, $S_1$ contains $\int_0 t^{-2p}dt$, displaying the coincident pointlike $X$ divergence. The uniform Gegenbauer bound $|C_n^p(\alpha)|\le C_n^p(1)=O(n^{2p-1})$, combined with the $O(n^{-2N-3})$ remainder, bounds the subtracted terms by $O(n^{2p-2N-4})$. Thus $N=\lceil p\rceil$ suffices for absolute convergence throughout $[-1,1)$.

For $|\alpha|<1$, a second parameter integral and
$\int_0^\infty du\,u^{\rho-1}K_\nu(u)=2^{\rho-2}\Gamma((\rho-\nu)/2)\Gamma((\rho+\nu)/2)$, with $\Re\rho>|\Re\nu|$, give
\begin{equation}\label{eq:AdS_Sodd_series}
 S_{2j+1}(\alpha)=\frac1{(2j)!\Gamma(p)}
 \sum_{m=0}^{\infty}\frac{2^{m-2}\alpha^m}{m!}g_m^{(2j)}(0),
\end{equation}
where
\begin{equation}
 g_m(\nu)=\Gamma\!\left(\frac{m+p-\nu}{2}\right)
           \Gamma\!\left(\frac{m+p+\nu}{2}\right),\qquad
 g_m^{(2j)}(0)=\left.\partial_\nu^{2j}g_m(\nu)\right|_{\nu=0}.
\end{equation}
The derivation uses $K_\nu(u)=\int_0^\infty e^{-u\cosh t}\cosh(\nu t)dt$. The series is geometrically convergent for $|\alpha|<1$, but not uniformly rapid near its endpoints; Eq.~\eqref{eq:AdS_Sodd_integral} is preferable there. Inserting these real sums in Eq.~\eqref{eq:AdS_X0_final} retains the upper-boundary phase $e^{+i\pi(2j+1)/2}$.

\section{Base-case series for dS detector integrals}
\label{app:dS_basecases}

\subsection{A common Gaussian-pole integral}

For $A>0$, $B\in\mathbb C$, and a pole $z$ off the real axis, set $\varsigma=\sgn\Im z$ and $\eta=\sqrt A\,z+iB/(2\sqrt A)$. Completing the Gaussian square gives
\begin{equation}
 \mathcal J_{A,B}(z):=\int_{\mathbb R}\frac{e^{-As^2-iBs}}{s-z}\,ds
 =\varsigma\,i\pi e^{-B^2/(4A)}w(\varsigma\eta),
 \qquad w(z):=e^{-z^2}\erfc(-iz).
 \label{eq:dS_gaussian_pole}
\end{equation}
Here $w$ is the Faddeeva function. For a real pole, $\varsigma=\pm1$ specifies its retained boundary side; it is determined before completing the square, not from $\Im\eta$. Analytic continuation makes Eq.~\eqref{eq:dS_gaussian_pole} valid for complex $B$. Integration by parts also gives
\begin{equation}
 \mathcal J'_{A,B}(z)=-(2Az+iB)\mathcal J_{A,B}(z)
 -2\sqrt{\pi A}\,e^{-B^2/(4A)}.
 \label{eq:dS_gaussian_pole_derivative}
\end{equation}
A prime differentiates the pole location. All sums over $n\in\mathbb Z$ below mean $\lim_{N\to\infty}\sum_{n=-N}^N$. Pairing opposite poles cancels their $1/n$ terms; the remaining $O(n^{-2})$ tails are algebraic and should be accelerated or bounded in numerical work.

\subsection{Excitation probability bases}

Use $A,B,\mathcal N_p$ from Eq.~\eqref{eq:dS_P_integral} and $z_n=-x+2\pi i n+i0$. The Mittag--Leffler expansions of $\csch[(s+x-i0)/2]$ and its square yield
\begin{align}
 P(\tfrac12;x)&=2\mathcal N_{1/2}\sum_{n\in\mathbb Z}(-1)^n\mathcal J_{A,B}(z_n),
 \label{eq:dS_P_base_half}\\
 P(1;x)&=4\mathcal N_1\sum_{n\in\mathbb Z}\mathcal J'_{A,B}(z_n).
 \label{eq:dS_P_base_one}
\end{align}
The residues in the first expansion are $2(-1)^n$; the double-pole coefficients in the second are $4$. In particular, the zero pole is above the contour. These expressions define the full real-$x$ family, including all derivatives needed in Eq.~\eqref{eq:dS_P_recursion}. Local uniform convergence of the paired meromorphic expansions, followed by Gaussian smearing, licenses finite-order shift differentiation.

\subsection{Integer correlation bases}

Let $\theta=\operatorname{arcosh}\kappa>0$. The pole decomposition
\begin{equation}
 K_1^\pm(s;\kappa)=\frac1{\sinh\theta}\sum_{n\in\mathbb Z}
 \left[\frac1{s+\theta-2\pi i n\pm i0}-\frac1{s-\theta-2\pi i n\pm i0}\right]
 \label{eq:dS_kernel_poles}
\end{equation}
fixes both residues and boundary signs. Define $\mathcal J^{\cos}_{A,B}(z)=[\mathcal J_{A,B}(z)+\mathcal J_{A,-B}(z)]/2$. Extending the even Gaussian--cosine test function to the full line gives the physical upper-boundary base
\begin{equation}
 X(1;\kappa)=-\frac{\beta_1}{\sinh\theta}
 \sum_{n\in\mathbb Z}\mathcal J^{\cos}_{\beta_2,\beta_3}(\theta+2\pi i n-i0).
 \label{eq:dS_X_base_one}
\end{equation}
The minus sign follows from the negative residue at $s=\theta$; at $n=0$ the pole in Eq.~\eqref{eq:dS_X_base_one} is below the contour. In contrast, the full-line lower-boundary coherence is
\begin{equation}
 C(1;\kappa)=\frac{\gamma_1}{\sinh\theta}\sum_{n\in\mathbb Z}
 \left[\mathcal J_{\gamma_2,\gamma_3}(-\theta+2\pi i n+i0)
 -\mathcal J_{\gamma_2,\gamma_3}(\theta+2\pi i n+i0)\right].
 \label{eq:dS_C_base_one}
\end{equation}
The coefficients in each base are evaluated at $p=1$. Equations~\eqref{eq:dS_X_base_one}--\eqref{eq:dS_C_base_one} retain complex lag dependence without conjugating the Gaussian factors.

\subsection{Half-integer correlation bases}

Put $a_\kappa=\sqrt{(\kappa-1)/(\kappa+1)}\in(0,1)$. Under $x=\tanh(s/2)$,
\begin{equation}
 ds=\frac{2\,dx}{1-x^2},\qquad
 \kappa-\cosh s=\frac{(\kappa-1)-(\kappa+1)x^2}{1-x^2}.
 \label{eq:dS_tanh_substitution}
\end{equation}
For brevity write $a=a_\kappa$ only within this subsection, and define the beta-function moments
\begin{align}
 E_n&=2\int_0^a\frac{x^{2n}\,dx}{\sqrt{1-x^2}\sqrt{a^2-x^2}}
 =a^{2n}\mathrm B(n+\tfrac12,\tfrac12)
 {}_2F_1(\tfrac12,n+\tfrac12;n+1;a^2),
 \label{eq:dS_C_Efinal}\\
 T_n&=2\int_a^1\frac{x^{2n}\,dx}{\sqrt{1-x^2}\sqrt{x^2-a^2}}
 =\pi a^{2n-1}{}_2F_1(\tfrac12-n,\tfrac12;1;-(1-a^2)/a^2),
 \label{eq:dS_X_tail_moment}\\
 O_n&=-2i\int_a^1\frac{x^{2n+1}\,dx}{\sqrt{1-x^2}\sqrt{x^2-a^2}}
 =-i\pi a^{2n}{}_2F_1(-n,\tfrac12;1;-(1-a^2)/a^2).
 \label{eq:dS_C_Ofinal}
\end{align}
Here $\mathrm B(u,v)=\Gamma(u)\Gamma(v)/\Gamma(u+v)$. The substitutions $x=a\sin u$ and $x^2=a^2+(1-a^2)\sin^2u$ evaluate the inner and outer integrals, respectively, with real square roots. In particular, the upper-boundary moment required for $X$ is
\begin{equation}
 I_{2n}^+(\kappa)=\int_0^1\frac{x^{2n}\,dx}
 {\sqrt{1-x^2}\sqrt{(\kappa-1)-(\kappa+1)x^2-i0}}
 =\frac{E_n+iT_n}{2\sqrt{\kappa+1}}.
 \label{eq:dS_X_base_half}
\end{equation}
Its imaginary part has the positive sign because the inverse square root on $x>a$ is $+i$ times the positive real inverse square root.

Expand the analytic test functions as
\begin{align}
 e^{-4\beta_2\operatorname{artanh}^2x}\cos[2\beta_3\operatorname{artanh}x]
 &=\sum_{n\ge0}b_{2n}x^{2n},\nonumber\\
 e^{-4\gamma_2\operatorname{artanh}^2x-2i\gamma_3\operatorname{artanh}x}
 &=\sum_{n\ge0}c_nx^n.
 \label{eq:dS_C_H}
\end{align}
Then, evaluating the Gaussian coefficients at $p=\tfrac12$,
\begin{align}
 X(\tfrac12;\kappa)&=\frac{\beta_1}{\sqrt{\kappa+1}}
 \sum_{n\ge0}b_{2n}(E_n+iT_n),
 \label{eq:dS_X_half_series}\\
 C(\tfrac12;\kappa)&=\frac{2\gamma_1}{\sqrt{\kappa+1}}
 \sum_{n\ge0}(c_{2n}E_n+c_{2n+1}O_n).
 \label{eq:dS_C_base_half}
\end{align}
For $C$, the two time halves add their even parts on $x<a$, whereas the opposite boundary phases isolate their odd parts on $x>a$.

These Taylor expansions admit termwise integration. Indeed, $\operatorname{artanh}z$ has bounded imaginary part in the unit disk, while its real part diverges logarithmically at $z=\pm1$. For positive Gaussian width coefficient, the negative square of that logarithm dominates every endpoint power, including those generated by any fixed derivative. The test functions thus extend smoothly to the closed disk, so their Taylor coefficients decrease faster than every inverse power. The moment weight is integrable at $a$ and 1. Finite-order $\kappa$ differentiation is justified distributionally: on compact subsets of $\kappa>1$, the differentiated polynomial-test-function seminorms grow at most polynomially with $n$, and the coefficient decay controls the sums. This establishes the higher-dimensional bases under Eqs.~\eqref{eq:dS_X_bootstrap}--\eqref{eq:dS_C_bootstrap}, without differentiating divergent ordinary integrals.

\subsection{An excitation series for numerical comparison}

For the $\mathrm{dS}_4$ figure, put $z=\Omega\sigma$, $h=2\pi\gamma/\sigma$, and $T(y)=1-\sqrt\pi(y/2)e^{y^2/4}\erfc(y/2)$. Splitting the spectrum~\eqref{eq:dS_spectral_recursion} at zero frequency and expanding its positive-frequency thermal factor gives
\begin{equation}
 P_1=\frac{\lambda^2e^{-z^2}}{4\pi}
 \left[T(2z)+\sum_{n=1}^\infty\{T(hn-2z)+T(hn+2z)\}\right].
 \label{eq:dS_P_erfc_series}
\end{equation}
Each term is a positive Gaussian integral, licensing the interchange by Tonelli's theorem. The unaccelerated remainder is $O(N^{-1})$. The large-argument expansion $T(y)\sim\sum_{j\ge1}(-1)^{j+1}(2j)!/(j!y^{2j})$ gives the tail of the bracketed sum after $n=N$ as
\begin{equation}
 R_N=\sum_{j=1}^{J}\frac{(-1)^{j+1}(2j)!}{j!h^{2j}}
 \left[\zeta(2j,N+1-2z/h)+\zeta(2j,N+1+2z/h)\right]
 +O(N^{-2J-1}),
 \label{eq:dS_P_erfc_tail}
\end{equation}
for fixed $z,h,J$ as $N\to\infty$; here $\zeta(s,a)$ is the Hurwitz zeta function. Fig. \ref{fig:dS4_probability_check} uses $N=64$, $J=8$, scaled erfc evaluation, and 12 asymptotic terms for $T(y)$ when $y\ge20$. Doubling $N$ and varying the time-contour displacement provide separate convergence checks. The reference quadrature uses $\delta=\gamma=1$, $|v|\le20\sigma$, and absolute and relative tolerances $2\times10^{-12}$; the reported smaller discrepancy is an observed comparison, not that tolerance as an error bound.

\section{Uniform BTZ endpoint asymptotics}
\label{app:BTZ_endpoint}
We prove Eq.~\eqref{eq:BTZ_smooth_log}; sharp release is the case
$\chi=1$. Define $G_{E,\zeta}(s)$ to be the bracket in
Eq.~\eqref{eq:BTZ_smooth_B}, and let $A_{n,\chi}$ denote
Eq.~\eqref{eq:BTZ_infall_An_start} with its integrand multiplied by
$\chi(\tau)\chi(\tau-s)$. The weighted zero image is
\begin{equation}
 \mathscr R^{(0)}_\chi(\tau;E)
 =\frac{\chi(\tau)^2}{4}
 -\frac{\chi(\tau)}{4\pi}
 \int_0^\tau ds\,\chi(\tau-s)G_{E,\zeta}(s),
 \qquad
 \mathscr R_\chi=\mathscr R^{(0)}_\chi+2\sum_{n\ge1}A_{n,\chi}.
 \label{eq:BTZ_weighted_images}
\end{equation}
The contact term is extracted once, as in
Sec.~\ref{sec:sharp_switching}. No derivative of $\chi$ occurs because
the profile is fixed as the integration endpoint moves.

Put $\varepsilon=\cos\tau$, $v=\sin\tau$,
$b_n=K_n-1=2\varrho^2\sinh^2(na)$, and $x_n=b_n\varepsilon$.
For one image, write $b=b_n$, $t=\tan(s/2)$, and
\begin{equation}
 A_b=2+b\varepsilon^2,\qquad
 t_+=\frac{\varepsilon[bv+\sqrt{b(b+2)}]}{A_b},\qquad
 t_-=\frac{b\varepsilon^2}{A_b t_+},\qquad
 D^-_n=\frac{A_b(t_+-t)(t+t_-)}{1+t^2}.
 \label{eq:BTZ_endpoint_factor}
\end{equation}
The substitution $t=-t_-+(t_++t_-)\sin^2 y$ removes the positive-branch
root, giving $ds/\sqrt{D^-_n}=4\,dy/(\sqrt{A_b}\sqrt{1+t^2})$.
For fixed $b$, its contribution tends to
$\chi_*^2[\tfrac12-\tfrac1{2\pi}\arctan\sqrt{2/b}]$.
The negative branch has the sign $-\sin(Es)$, as specified in
Sec.~\ref{subsec:BTZ_infall_rate}; together with the reflected term it
tends to the integral in Eq.~\eqref{eq:BTZ_smooth_B}.

For summation, these limits must be uniform. Fix a sufficiently small
$x_0>0$. For $b\ge b_1>0$ and $x=b\varepsilon\le x_0$, the positive
root lies at $s=O(x)$. The preceding substitution, with
$\chi(\tau)\chi(\tau-s)=\chi_*^2+O(\varepsilon+s)$, gives a uniform
positive-branch error $O(x+\varepsilon)$. On the negative branch,
the region $s=O(x)$ contributes $O(x)$ because $\sin(Es)=O(s)$.
For $s$ beyond a sufficiently large multiple of $x$, the difference
from the limiting inverse square root is bounded by
$C(x/s^2+x\varepsilon/s^3)$. Multiplication by $\sin(Es)$ and
integration give $O(x[1+|\log x|]+\varepsilon)$.
The reflected denominator is at least one, so its error is
$O(x+\varepsilon)$. Hence, uniformly in this image band,
\begin{equation}
 A_{n,\chi}=B_{\chi,\zeta}
 -\frac{\chi_*^2}{2\pi}\arctan\sqrt{\frac2{b_n}}
 +O\!\left(x_n[1+|\log x_n|]+\varepsilon\right).
 \label{eq:BTZ_uniform_image}
\end{equation}
In particular, the fixed-image limit differs from $B_{\chi,\zeta}$.
Its deficit is exponentially summable. Since
$b_{n+1}/b_n>e^{2a}$, the errors in Eq.~\eqref{eq:BTZ_uniform_image}
have a bounded sum over the
\begin{equation}
 N(\varepsilon)=
 \left\lfloor\frac1a\operatorname{arsinh}
 \sqrt{\frac{x_0}{2\varrho^2\varepsilon}}\right\rfloor
 =\frac{1}{2a}\log\frac1\varepsilon+O(1)
 \label{eq:BTZ_image_count}
\end{equation}
images in this band. The intermediate band $x_0<x_n<2$ contains
boundedly many images with uniformly integrable simple roots.
For $x_n\ge2$ and $\tau$ sufficiently near $T$,
$D^-_n\ge(x_n/4)\sin s$ on $0\le s\le\tau$, so
$A_{n,\chi}=O(x_n^{-1/2})$ and the remaining tail is uniformly
summable. The bounded zero image thus leaves
$\mathscr R_\chi=2N B_{\chi,\zeta}+O(1)$.
Using $\varepsilon=\sin\delta$ proves Eq.~\eqref{eq:BTZ_smooth_log}.

For positivity under monotone onset, put $w(s)=\chi(T-s)$.
For $\zeta=\pm1$,
$G_{E,\zeta}(s)=2\sin[(E+\zeta/2)s]/\sin s$.
For $k>0$ and $0<L\le T$, integration against the primitive
$(1-\cos ks)/k$ first gives
$J_L(k):=\int_0^L\sin(ks)/\sin s\,ds>0$.
With $f(s)=s/\sin s$, a second integration by parts gives
\begin{equation}
 J_L(k)=f(L)\operatorname{Si}(kL)
 -\int_0^L f'(s)\operatorname{Si}(ks)\,ds
 \le\frac\pi2\operatorname{Si}(\pi),
 \qquad
 B_{\chi,\zeta}(E)\ge
 \chi_*^2\left[\frac12-\frac{\operatorname{Si}(\pi)}4\right]>0.
 \label{eq:BTZ_positive_coefficient}
\end{equation}
Here $0<\operatorname{Si}(x)\le\operatorname{Si}(\pi)<2$ for $x>0$.
To obtain the bound for the weighted integral, write
$w(s)=w(T)+\int_s^T[-w'(L)]\,dL$ and use oddness in $k$.
The total nonnegative weight is $w(0)=\chi_*$.
The transparent coefficient is the average of the two reflecting
coefficients and obeys the same bound.

Finally, a switch-on whose extension by zero is $C^\infty$ is flat at
release. When a null-image root crosses that endpoint, a smooth root
coordinate pairs the inverse-square-root boundary distribution with a
smooth weight flat at the endpoint; its contribution is smooth in
$\tau$. The readout endpoint is regular for each nonzero image since
$D^-_n(0)=b_n\cos^2\tau>0$ for $\tau<T$. On compact subintervals of
$(0,T)$ the image tails and their derivatives converge geometrically.
Thus smoothing the onset removes the finite-time glitches without
altering the endpoint asymptotic law.

\subsection{Numerical convergence details}
\label{app:BTZ_numerical_details}
\begin{figure}[!htbp]
 \centering
 \includegraphics[width=0.48\textwidth]{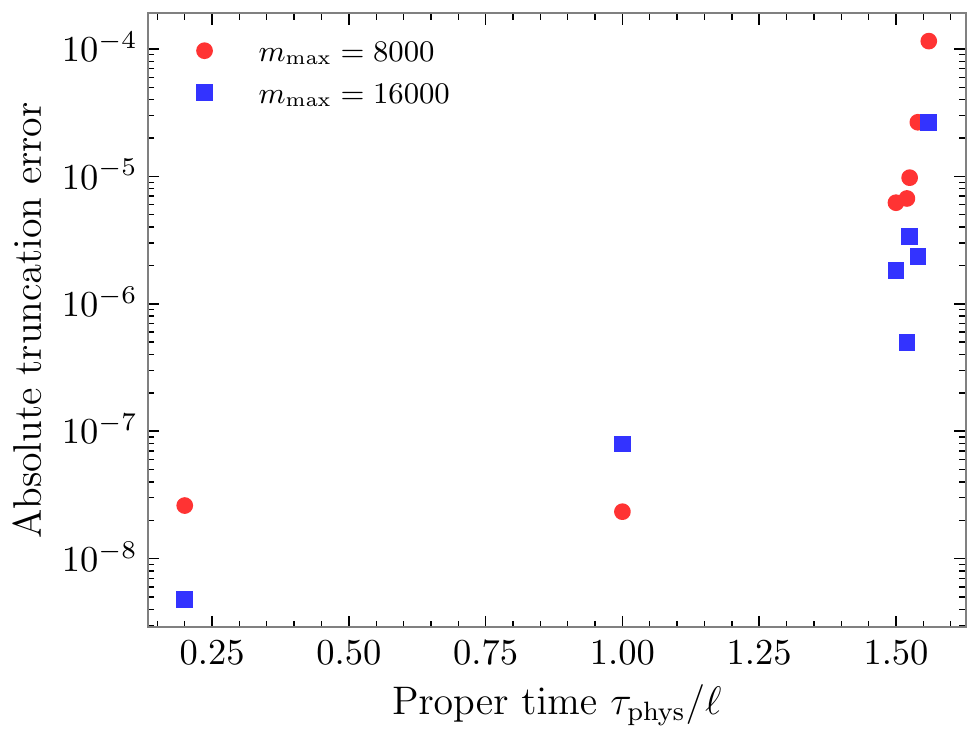}
 \caption{Absolute mode-truncation error against image-integral
 quadrature, with the parameters of Fig.~\ref{fig:BTZ_infall_check}
 and $n_{\max}=600$. The error need not decrease at every point when $m_{\max}$ is doubled. Only seven times were checked; these data give no uniform bound at a glitch or the singularity.}
 \label{fig:BTZ_convergence}
\end{figure}

% Moved from the main-text BTZ numerical discussion; data and tolerances unchanged.
The displayed curve in Fig.~\ref{fig:BTZ_infall_check} uses shape-preserving cubic interpolation of 464 evaluated series points, clustered near both visible glitches; independent midpoint checks bound the observed interpolation discrepancy by $10^{-4}$. The 33 quadrature markers are evaluated independently. The image cutoff $n_{\max}=600$ versus $800$ changes the checked rates by at most $7.4\times10^{-10}$, whereas the mode truncation is larger and oscillatory, as Fig.~\ref{fig:BTZ_convergence} shows. At an exact glitch the generic mode tail is only $O(m_{\max}^{-1/2})$.

For Fig.~\ref{fig:BTZ_endpoint_comparison}, $E=0$ and $\zeta=0$, so the negative direct branch has zero real part. We integrate the positive branch using Eq.~\eqref{eq:BTZ_endpoint_factor} to remove the square root. The zero image contributes $\chi(\tau)^2/4$. We use 160 positive images and check the image cutoff and quadrature separately. The logarithmic coefficient is not fitted.

\clearpage
\section{Integral-to-analytic lookup}
\label{app:lookup}
The prefactors and time variables in Table~\ref{tab:analytic_lookup} follow the cited equations. For static AdS and dS, set $f_P(s)=e^{-\gamma^2s^2/(4\sigma^2)-i\gamma\Omega s}$, $f_X(s)=e^{-As^2}\cos(B_Xs)$, and $f_C(s)=e^{-As^2-iB_Cs}$. Their coefficients are given in Appendix~\ref{app:coefficients}. The AdS kernel uses the boundary value of $F_p$ in Eq.~\eqref{eq:AdS_branch}, including both direct and boundary terms. The dS kernels are $K_p^\pm$. For Minkowski, $f_M(s)=e^{-s^2/(4\sigma^2)}$ and $a,b$ are defined in Eq.~\eqref{eq:Mink_C_parameters}. We require $\sigma>0$. Dimension raising holds the formal coupling fixed; physical couplings that differ between dimensions supply an additional ratio.

\begingroup
\setlength{\tabcolsep}{3pt}
\renewcommand{\arraystretch}{1.17}
\begin{table}[!ht]
\centering
\caption{Detector integrals, analytic representations, and domains, ordered as in the main text. ``Abel'' specifies a boundary limit, not ordinary convergence of an unsubtracted series.}\label{tab:analytic_lookup}
\begin{tabular}{@{}p{0.11\textwidth}p{0.07\textwidth}p{0.29\textwidth}p{0.29\textwidth}p{0.18\textwidth}@{}}
\toprule
\raggedright Geometry, state & \raggedright Term & \raggedright Starting integral & \raggedright Analytic reduction & \raggedright Domain or qualification \tabularnewline
\midrule
\raggedright Minkowski vacuum & \raggedright $P$ & \raggedright $\int_{\mathbb R}f_M(s)e^{-i\Omega s}W(s)ds$; Eq.~\eqref{eq:PD_general} & \raggedright Positive spectral integral, Eq.~\eqref{eq:Mink_P_positive}; Kummer or cylinder function, Eqs.~\eqref{eq:excitationMinkclosedform}, \eqref{eq:Mink_P_cylinder} & \raggedright $\mathcal D\ge3$; all real signed frequencies \tabularnewline
\raggedright  & \raggedright $X$ & \raggedright $\int_0^\infty f_M(s)[L^2-(s+i0)^2]^{-p}ds$; Eq.~\eqref{eq:Mink_X_halfline} & \raggedright Upper-lip Tricomi function, Eq.~\eqref{eq:Mink_X_closed} & \raggedright $L>0$, $t_0=0$ for the closed form \tabularnewline
\raggedright  & \raggedright $C$ & \raggedright $\int_{\mathbb R}e^{-as^2-ibs}[L^2-(s-i0)^2]^{-p}ds$; Eq.~\eqref{eq:Mink_C_integral} & \raggedright Bessel--Gaussian master, Eq.~\eqref{eq:Mink_C_master}; error-function $p=1$ base and raising, Eqs.~\eqref{eq:Mink_C_one}, \eqref{eq:Mink_C_raising} & \raggedright Restore Eq.~\eqref{eq:Mink_C_physical}; complex $b$ uses the holomorphic difference \tabularnewline
\raggedright Global AdS vacuum & \raggedright $P$ & \raggedright $\int_{\mathbb R} f_P(s)W(s)\,ds$; Eq.~\eqref{eq:AdS_P_after_pullback} & \raggedright Gaussian-damped Gegenbauer sum, Eq.~\eqref{eq:AdSexcitation} & \raggedright $\mathcal D\ge3$; finite static radius \tabularnewline
\raggedright  & \raggedright $X$ & \raggedright $\int_0^\infty f_X(s)F_p(-s-i0,x)\,ds$; Eq.~\eqref{eq:AdS_X_integral} & \raggedright Subtracted half-Fourier sum and restored Abel moments, Eqs.~\eqref{eq:AdS_X0_final}, \eqref{eq:AdS_Sodd_integral} & \raggedright Distinct spatial positions; $x\in[-1,1)$ \tabularnewline
\raggedright  & \raggedright $C$ & \raggedright Full two-detector integral, Eq.~\eqref{eq:C_general} & \raggedright Separate Gaussian mode transforms, Eq.~\eqref{eq:AdS_C_final} & \raggedright Includes coincidence; $C=P$ for identical simultaneous profiles \tabularnewline
\raggedright dS Bunch--Davies & \raggedright $P$ & \raggedright $\int_{\mathbb R}f_P(s)\csch^{2p}[(s+x-i0)/2]ds$; Eq.~\eqref{eq:dS_P_integral} & \raggedright Pole bases, Eqs.~\eqref{eq:dS_P_base_half}--\eqref{eq:dS_P_base_one}; normalized recurrence, Eq.~\eqref{eq:dS_P_recursion} & \raggedright $R<\alpha$; take shift derivatives before $x=0$ \tabularnewline
\raggedright  & \raggedright $X$ & \raggedright $\int_0^\infty f_X(s)K_p^+(s;\kappa)ds$; Eq.~\eqref{eq:dS_X_integral} & \raggedright Gaussian-pole or split-moment base, Eqs.~\eqref{eq:dS_X_base_one}, \eqref{eq:dS_X_half_series}; Eq.~\eqref{eq:dS_X_bootstrap} & \raggedright $\kappa>1$; other parameters fixed under differentiation \tabularnewline
\raggedright  & \raggedright $C$ & \raggedright $\int_{\mathbb R} f_C(s)K_p^-(s;\kappa)ds$; Eq.~\eqref{eq:dS_C_integral} & \raggedright Even/odd moment or pole base, Eqs.~\eqref{eq:dS_C_base_one}, \eqref{eq:dS_C_base_half}; Eq.~\eqref{eq:dS_C_bootstrap} & \raggedright $\kappa>1$ in these bases; coincidence from the original full-line integral \tabularnewline
\raggedright BTZ Hartle--Hawking & \raggedright $\dot{\mathcal F}$ & \raggedright Sharp image integrals $A_n$, Eq.~\eqref{eq:BTZ_infall_An_start} & \raggedright Integrated Legendre modes and regular zero-image integral, Eqs.~\eqref{eq:BTZ_infall_An_double_series}, \eqref{eq:BTZ_infall_n0_term} & \raggedright $\mathcal D=3$; release switch-on; $0<\tau<\pi/2$ \tabularnewline
\bottomrule
\end{tabular}
\end{table}
\endgroup

\bibliographystyle{apsrev4-2}
% Select the bibliography beside the uploaded PRX_final assets when present.
\IfFileExists{PRX_final/refs.bib}{\bibliography{PRX_final/refs}}{\bibliography{refs}}
\end{document}